\documentclass[aps,prl,twocolumn,longbibliography,superscriptaddress,tightenlines]{revtex4-1}

\newcommand{\be}{\begin{equation}}
\newcommand{\ee}{\end{equation}}
\newcommand{\bea}{\setlength\arraycolsep{2pt} \begin{eqnarray}}
\newcommand{\eea}{\end{eqnarray}}
\newcommand{\nn}{\nonumber}

\def\0{{\sst{(0)}}}
\def\1{{\sst{(1)}}}
\def\2{{\sst{(2)}}}
\def\3{{\sst{(3)}}}
\def\4{{\sst{(4)}}}
\def\5{{\sst{(5)}}}
\def\6{{\sst{(6)}}}
\def\7{{\sst{(7)}}}
\def\8{{\sst{(8)}}}
\def\sst#1{{\scriptscriptstyle #1}}

\usepackage{multirow}
\usepackage{xcolor}
\usepackage{bm}
\usepackage{graphicx}
\usepackage{epsfig,psfrag}
\usepackage{amsmath}
\usepackage{amssymb}
\usepackage{amsbsy}
\usepackage{amsthm}
\usepackage{amsfonts}
\usepackage{wasysym}
\usepackage{bbm}
\usepackage{tabularx}
\usepackage{euscript}
\usepackage{enumerate}
\usepackage{amsfonts}
\usepackage{exscale}
\usepackage{bbold}
\usepackage{float}
\usepackage{slashed}
\usepackage{subfigure}
\usepackage{comment}
\usepackage[]{hyperref}
\usepackage[]{natbib}

\def\bea{\begin{eqnarray}}
\def\eea{\end{eqnarray}}
\def\nn{\nonumber}
\def\ba{\begin{array}}
	\def\ea{\end{array}}
\def\nn{\nonumber}

\def\sgn{\text{sgn}}

\def\kc#1{\left(#1\right)}
\def\kd#1{\left[#1\right]}

\def\sgn{{\rm sgn}}

\def\be{\begin{equation}}       \def\ee{\end{equation}}
\def\bea{\begin{eqnarray}}      \def\eea{\end{eqnarray}}
\def\ba{\begin{array}}
	\def\ea{\end{array}}
\def\bnum{\begin{enumerate} }
	\def\enum{\end{enumerate}}

\def\nn{\nonumber}

\def\=>{\Rightarrow}
\def\>{\rightarrow}

\def\eye2{Fathbb{I}}

\DeclareMathOperator\csch{csch}

\def\Teta{\hat{\mathfrak T}_\eta}
\def\MK{\hat{\mathfrak M}_K}
\def\Mh{\hat{\mathfrak M}_h}

\begin{document}

\title{Hall viscosity and hydrodynamic inverse Nernst effect in graphene}

\author{Zhuo-Yu~Xian}
\affiliation{Institute for Theoretical Physics and Astrophysics and W{\"u}rzburg-Dresden Cluster of Excellence ct.qmat,\\
	Julius-Maximilians-Universit{\"a}t W{\"u}rzburg, D-97074 W{\"u}rzburg, Germany}

\author{Sven~Danz}
\affiliation{Institute for Theoretical Physics and Astrophysics and W{\"u}rzburg-Dresden Cluster of Excellence ct.qmat,\\
	Julius-Maximilians-Universit{\"a}t W{\"u}rzburg, D-97074 W{\"u}rzburg, Germany}
\affiliation{Peter Gr{\"u}nberg Institute - Quantum Computing Analytics (PGI 12), Forschungszentrum J{\"u}lich, D-52425 J{\"u}lich, Germany}
\affiliation{Theoretical Physics,
Saarland University, D-66123 Saarbr{\"u}cken, Germany}
	
\author{David~Rodr\'iguez~Fern\'andez}
\affiliation{ Instituut-Lorentz for Theoretical Physics, Universiteit Leiden, P. O. Box 9506, 2300 RA Leiden, The Netherlands}
\affiliation{Institute for Theoretical Physics and Astrophysics and W{\"u}rzburg-Dresden Cluster of Excellence ct.qmat,\\
	Julius-Maximilians-Universit{\"a}t W{\"u}rzburg, D-97074 W{\"u}rzburg, Germany}
	
\author{Ioannis~Matthaiakakis}
\affiliation{Dipartimento di Fisica, Universit\`a di Genova and I.N.F.N. - Sezione di Genova, \\ via Dodecaneso 33, I-16146, Genova, Italy}
\affiliation{Institute for Theoretical Physics and Astrophysics and W{\"u}rzburg-Dresden Cluster of Excellence ct.qmat,\\
	Julius-Maximilians-Universit{\"a}t W{\"u}rzburg, D-97074 W{\"u}rzburg, Germany}

\author{Christian~Tutschku}
\affiliation{Institute for Theoretical Physics and Astrophysics and W{\"u}rzburg-Dresden Cluster of Excellence ct.qmat,\\
	Julius-Maximilians-Universit{\"a}t W{\"u}rzburg, D-97074 W{\"u}rzburg, Germany}
\affiliation{Fraunhofer IAO, Fraunhofer Institute for Industrial Engineering IAO, D-70569 Stuttgart, Germany}

\author{Raffael~L.~Klees}
\affiliation{Institute for Theoretical Physics and Astrophysics and W{\"u}rzburg-Dresden Cluster of Excellence ct.qmat,\\
	Julius-Maximilians-Universit{\"a}t W{\"u}rzburg, D-97074 W{\"u}rzburg, Germany}

\author{Johanna~Erdmenger}
\affiliation{Institute for Theoretical Physics and Astrophysics and W{\"u}rzburg-Dresden Cluster of Excellence ct.qmat,\\
	Julius-Maximilians-Universit{\"a}t W{\"u}rzburg, D-97074 W{\"u}rzburg, Germany}

\author{Ren\'e~Meyer}
\affiliation{Institute for Theoretical Physics and Astrophysics and W{\"u}rzburg-Dresden Cluster of Excellence ct.qmat,\\
	Julius-Maximilians-Universit{\"a}t W{\"u}rzburg, D-97074 W{\"u}rzburg, Germany}

\author{Ewelina~M.~Hankiewicz}
\affiliation{Institute for Theoretical Physics and Astrophysics and W{\"u}rzburg-Dresden Cluster of Excellence ct.qmat,\\
	Julius-Maximilians-Universit{\"a}t W{\"u}rzburg, D-97074 W{\"u}rzburg, Germany}

\begin{abstract}
Motivated by Hall viscosity measurements in graphene sheets, we study hydrodynamic transport of electrons in a channel of finite width in external electric and magnetic fields. We consider electric charge densities varying from close to the Dirac point up to the Fermi liquid regime.
We find two competing contributions to the hydrodynamic Hall and inverse Nernst signals that originate from the Hall viscous and Lorentz forces. 
This competition leads to a non-linear dependence of the full signals on the magnetic field and even a cancellation at different critical field values for both signals. In particular, the hydrodynamic inverse Nernst signal in the Fermi liquid regime is dominated by the Hall viscous contribution.
We further show that a finite channel width leads to a suppression of the Lorenz ratio, while the magnetic field enhances this ratio. All of these effects are predicted in parameter regimes accessible in experiments.
\end{abstract}
\date{\today}

\maketitle

\textit{Introduction.}\textemdash In the last two decades, electronic fluids have become a main research object in condensed matter physics, allowing for the realization of new transport effects. The main platform for electron hydrodynamics studies and applications is graphene \cite{science:aac8385,Crossno:LorentzRatio,Sulpizio:2019NaturePoiseuille,Ku:2020NatureViscousFlow,Jenkins:2020Imaging,Vool2021,Sulpizio:2019NaturePoiseuille,Ku:2020NatureViscousFlow,Geim:2016Negative,Crossno:LorentzRatio,Gallagher:2019MontumRelaxation,Berdyugin:2019SciHall,ella2019simultaneous,bandurin2018fluidity}, 
for which hydrodynamic behavior can be experimentally accessed both close to the Dirac point as well as in the Fermi liquid regime  \cite{Lucas:2018JPCM,Narozhny:2017HydroGraphene,Narozhny2022}. 
In particular, graphene  allows for the investigation of new transport effects induced by the Hall viscosity $\eta_H$ that was recently measured to be of the same magnitude as the shear viscosity $\eta$ \cite{Berdyugin:2019SciHall} and, hence, strongly affects graphene's hydrodynamic properties. 
More precisely, both $\eta$ and $\eta_H$ determine the transport properties of electronic fluids in finite sample geometries  \cite{Sulpizio:2019NaturePoiseuille,Ku:2020NatureViscousFlow,Geim:2015PRBNonlocalHydro}. 
This was observed experimentally in GaAs \cite{Molenkamp:PRBGaAs,Jong:PRBGaAs,gusev2018viscous},
graphene
\cite{Ku:2020NatureViscousFlow,Sulpizio:2019NaturePoiseuille,Geim:2016Negative,Berdyugin:2019SciHall}, and PdCoO$_2$ compounds \cite{science:aac8385}. 
Moreover, the full ballistic-to-hydrodynamic crossover \cite{Molenkamp:PRBGaAs,Jong:PRBGaAs}
as well as a negative magnetoresistance due to  viscous effects \cite{PhysRevLett.117.166601,Alekseev:2015PRL,Alekseev:magnetoresistance,Alekseev:2017PRBMagnetoresistance} were observed in channel geometries.

The Hall viscosity $\eta_H$ breaks parity and time reversal symmetries \cite{Avron:1995PRLHall,Jensen:2011xb,Hoyos:2014Hall},
and is generated in a parity-invariant electronic fluid by an external  magnetic field \cite{Alekseev:2015PRL,Berdyugin:2019SciHall}. 
Based on this, in \cite{Matthaiakakis:2020PRB},  a \textit{non-monotonic} magnetic field dependence of the Hall effect in GaAs Fermi liquids was predicted. This arises from the competition of the Hall viscous and Lorentz forces acting on the fluid. %
In contrast to GaAs, in graphene, the relativistic Dirac spectrum allows for new hydrodynamic transport effects  close to the Dirac point. 

In this Letter, we find new contributions to the  thermoelectric transport of the electronic fluid in graphene in the presence of an external magnetic field. Similarly to the Hall effect in \cite{Matthaiakakis:2020PRB}, we show that the hydrodynamic inverse Nernst effect receives, in addition to the contribution from the Lorentz force, a new contribution from the Hall viscosity. This  contribution arises both close to charge neutrality as well as in the Fermi liquid regime. Moreover, close to the Dirac point \cite{Footnote1}, we find that the 
quantum critical (or incoherent) conductivity $\sigma_Q$  \cite{Markus:2008quantumcritical,Lucas:2018JPCM}, originating from momentum-conserving scattering between electrons and holes, crucially contributes to the hydrodynamic inverse Nernst signal as well. In addition, we predict a cancellation between the Hall viscous and Lorentz force contributions in both the hydrodynamic Hall and inverse Nernst signals at different critical magnetic fields of the order of $10~{\rm mT}$. The critical fields increase with increasing $\mu/k_B T$.
Let us emphasize that our new hydrodynamic inverse Nernst effect scales very differently with a system size and an out-plane-magnetic field in comparison 
with Nernst/inverse Nernst effects in ballistic or diffusive regimes of metals, ferromagnets and spin-orbit systems \cite{Behnia_2016,Sakai_2018,Rothe2012}.

Finally, we calculate the thermal and electric conductivities for our flows and show their ratio violates the Wiedemann-Franz law \cite{Franz:1853WFLaw,ashcroft1976,Crossno:LorentzRatio} and, for the first time, show that the magnitude of the violation is a monotonically increasing function of both the external magnetic field and the width of the channel.


\begin{figure}
	\includegraphics[width=0.35\textwidth]{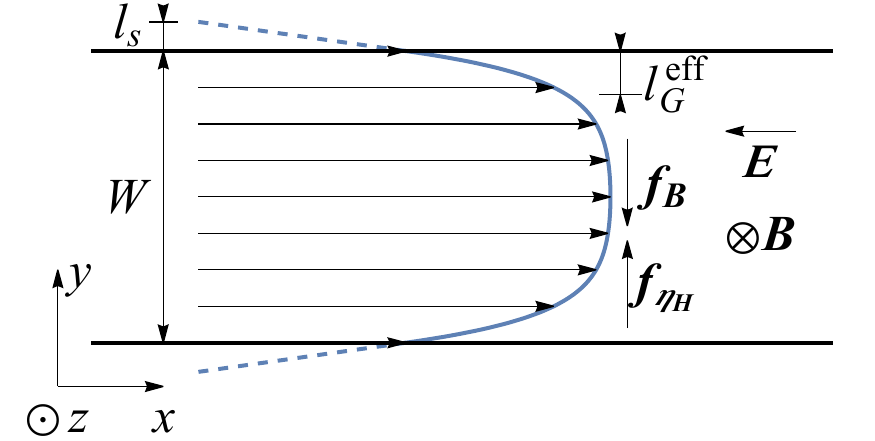}
	\caption{The velocity profile of an electron fluid in a graphene channel of width $W$ in the $x$-$y$-plane. The electric field $\bm E$ is applied along the channel, while the magnetic field $\bm B$ is applied perpendicular to the channel. Furthermore, the slip length $l_s$ as well as the effective Gurzhi length $l_G^{\rm eff}$ are depicted. The Lorentz force density ${\bm f}_B$ and Hall viscous force density ${\bm f}_{\eta_H}$ act on the fluid in opposite directions across the channel.
	}
	\label{fig:Cartoon}
\end{figure}

\textit{Hydrodynamics in a graphene channel.}\textemdash
We consider a graphene channel of finite width $W$ and length $L\gg W$ subjected to an electric field ${\bm E}=E_x{\bm e}_x$ and a magnetic field ${\bm B}=B_z{\bm e}_z$, as shown in Fig.~\ref{fig:Cartoon}. 
The electrons, with charge $-e$, are pumped through the channel by ${\bm E}$ and are deflected towards the channel boundaries by the Lorentz force density ${\bm f}_B$ and the Hall viscous force density ${\bm f}_{\eta_H}$ defined in Eq.~\eqref{forcedef} below.
The Hall viscous force density ${\bm f}_{\eta_H}$ is induced by the Hall viscosity $\eta_H$, defined in the Supplemental Material (SM) \cite{supMat}, which is a parity-breaking dissipationless transport coefficient in the hydrodynamic expansion of the stress tensor \cite{Avron:1995PRLHall,avron1998odd,Jensen:2011xb}.
These forces trigger the transverse temperature gradient $\Delta T$ (inverse Nernst effect) and Hall voltage $\Delta V$ across the channel
\cite{Footnote2}. To avoid excessive Joule heating and to stay in the linear response regime, we consider small electric fields. 
We also restrict to $|B_z| \ll 1~{\rm T}$ to avoid the formation of Landau levels.

For the validity of electron hydrodynamics in the graphene channel, the electron fluid should reach local equilibrium via fast scattering compared to other effects, such as momentum relaxation, energy relaxation, and the effect of a finite channel width. 
We maintain the hierarchy between the corresponding characteristic times by considering reasonable parameters, such as $\mu / k_B T > 0.1$ and $100~{\rm K} < T < 300~{\rm K}$, as explained in the SM \cite{supMat} \cite{Footnote3}. 
Within this region, the electron-electron scattering, which is characterized by the dimensionless coupling $\alpha \approx 0.5$, leads to the electron-electron scattering time $\tau_{ee} \sim 0.1~{\rm ps}$ \cite{Lucas:2018JPCM}.
The momentum relaxation time $\tau_{\rm MR}$ is of order of magnitude of $1~{\rm ps}$ for our range of parameters, while \cite{Markus:2008quantumcritical,Gallagher:2019MontumRelaxation,Hwang:2008MomentumPhonon,Efetov:PRL:PhononExp,Shaffique:PNAS:self-consistent}. 
the energy relaxation time $\tau_\text{ER}$ is of order of magnitude of $100~{\rm ps}$ \cite{Cooling:Bistritzer:PRL,Cooling:Tse:PRB,Narozhny:2021energyrelaxation,Disorder-Electron-Phonon,Graham:2013Supercollisions,Betz:2013NatPh:SupercollisionExperiment,Fong:PRX:EnergyRelaxationExperiment}.
Finally, the ``ballistic'' time $\tau_B \sim W/v_F$ where $W$ is the width of the channel and $v_F \simeq 10^6 {\rm m/s}$ graphene's Fermi velocity \cite{Alekseev:magnetoresistance}. 
To avoid ballistic effects, we consider a width of the order $W \sim 1~{\rm \mu m}$ for which $\tau_B \sim 1~{\rm ps}$. 
All of the above considerations show, that for the range of parameters we consider i) we are always in the hydrodynamic regime and ii) momentum relaxation must be taken into account, while energy relaxation can be neglected, since $\tau_\text{ER}$ is much larger than all other time scales. Neglecting energy relaxation does not lead to qualitative changes of our results. 

The hydrodynamic variables describing the fluid are the flow velocity $\bm{v}$, the electrochemical potential $\mu_\text{tot}$, and the total temperature $T_\text{tot}$. 
At global equilibrium and vanishing external sources, these are the usual velocity, chemical potential, and temperature characterizing a thermal state.
This changes when external sources are turned on. 
In particular, within linear response, a non-zero electric field $E_x$ and temperature gradient $\nabla_x T$ can be incorporated into $\mu_{\rm tot}$ and $T_{\rm tot}$. Thus, restricting ourselves to small fluctuations around global equilibrium with zero velocity and constant chemical potential and temperature, we consider the following ansatz for our hydrodynamic variables
\begin{subequations}\label{Eq:Ansatz}\begin{align}
\bm{v} &= (v_x(y), v_y(y))~, \\
\mu_\text{tot} &= \mu + \delta\mu(y) + e x E_x~, \\ 
T_\text{tot} &= T + \delta T(y) + x\nabla_x T~,
\end{align}\end{subequations}
where $\mu$ and $T$ are the equilibrium chemical potential and temperature respectively, whereas $\delta \mu, ~\delta T$ are the fluctuation fields. 
Linear response requires $\delta\mu\ll \mu $ and $ \delta T\ll T$, while the $x$-translation invariance along the channel implies the fluctuation fields can depend only on $y$. 
Note that $\delta \mu$ can be further split into two contributions as $\delta\mu(y)=\delta\mu_T(y)-e\phi_V(y)$, where $\delta \mu_T$ is due to deviation from thermal equilibrium and $\phi_V$ is the Vlasov potential generated by the backreaction of the fluid to the electric field \cite{Narozhny:2019ehg,Alekseev:2015PRL,Narozhny:2017HydroGraphene}.

Due to graphene's relativistic dispersion relation, the hydrodynamic variables are fixed by the equations of relativistic hydrodynamics at linear order in the velocities \cite{landau2013fluid,Hartnoll:2007Nernst,Jensen:2011xb} \cite{Footnote4}, with a limit velocity  $v_F$ and an effective electrical field $(c/v_F)\bm E$. 
Using the ansatz in Eq.~\eqref{Eq:Ansatz}, the hydrodynamic equations read \cite{supMat}
\begin{subequations}
\label{Eq:ODE}
\begin{align}
	w\partial_yv_y = wv_y' &= 0, \label{ODE1}
	\\
	  e n E_1-\eta v_x'' - {\bm f_{\eta_H}}_{,x}  &= -\frac{P_x}{\tau_\text{MR} v_F^2} + {\bm f_B}_{,x}  , \label{ODE2} 
	\\
	\delta p' -{\bm f_{\eta_H}}_{,y} -\eta v_y'' - e n \phi_V'&= -\frac{P_y}{\tau_\text{MR} v_F^2} + {\bm f_B}_{,y} , \label{ODE3}
	\\
	\partial_yJ_y &= 0, \label{ODE4}\\ \label{forcedef}
	{\bm f_B} =\frac{1}{n}\epsilon^{ij} {\bm e}_i J_j B_z 
	\,,\quad & {\bm f_{\eta_H}}=\frac{ \eta_H}{n} \epsilon^{ij} {\bm e}_i v''_j \,.
\end{align}
\end{subequations}
Equations \eqref{ODE1}\textendash\eqref{ODE4} correspond to the conservation of energy, longitudinal momentum, transversal momentum, and charge, respectively. The transport coefficients
$\eta$ and $\eta_H$ are the shear and Hall viscosities, which are both positive in our setup as shown in Fig.~S2 of the SM \cite{supMat}.
Moreover, $n$, $n_E$, $p$, $w$, and $s$ are the equilibrium particle number density, energy density, pressure,
enthalpy, and entropy density, respectively. 
They are all known functions of $\mu$ and $T$ satisfying $w = n_E + p=3n_E/2=\mu_Tn+sT$ \cite{supMat}. 
The deviation of the pressure away from equilibrium is related to the hydrodynamic variables by $\delta p = n \, \delta\mu_T + s\, \delta T$. 
A central ingredient of our analysis is that the Lorentz force density ${\bm f}_B$ and the Hall viscous force density ${\bm f}_{\eta_H}$ have opposite signs. 
The relative and overall signs of the force densities stem from the following considerations: The origin of the relative sign is the opposite signs of the Poiseuille flow, $v_x>0$, and its curvature, $v_x''(y)<0$, in the coordinate system of Fig.~\ref{fig:Cartoon}. 
The overall sign is set by $e B_z$, which enters both force densities in the same way. 
The forces point in exactly opposite directions since the dominant $v_x$ and $v_x''$ generate ${\bm f}_B$ and ${\bm f}_{\eta_H}$, respectively. 
These forces are in turn responsible for creating the velocity profile $v_y$, whose magnitude is much smaller than that of $v_x$,  $|v_x|\gg |v_y|$.  
Summarizing, the antiparallel configuration of the forces gives rise to the unconventional thermoelectric response we find below.

The charge currents $J_{x,y}$ and momenta $P_{x,y}$ entering the conservation equations \eqref{Eq:ODE} are
\begin{subequations}\label{Currents}
\begin{align}
	J_x &= -env_x+\sigma_Q\kc{B_zv_y+ E_2},	\\
	J_y &= -env_y+\sigma_Q\kd{-B_z v_x +\frac1e \kc{\delta\mu'-\frac{\mu}{T}\delta T'}},\\
	P_x &= w v_x, \qquad P_y = w v_y.
\end{align}
\end{subequations}
To simplify notation, we have recombined the sources into $E_1 = E_x + s \nabla_x T/(e n)$ and $E_2 = E_x - \mu \nabla_xT/(eT)$. From Eq.~\eqref{Currents}, we see that $E_1$ drives the momentum density $P_x$ and $E_2$ the quantum critical current density $J_Q = \sigma_Q E_2$.
In principle, we should also consider the electrostatic Poisson equation to calculate the Vlasov field and close our system of equations.
However, since the problem is linear, we can combine $\phi_V'$ and $\delta\mu_T'$ to $\delta \mu'$ in Eq.~\eqref{ODE3} and solve Eq.~\eqref{Eq:ODE} for the variables $v_x$, $v_y$, $\delta\mu$, and $\delta T$. 
Note that the above considerations are valid for an electron-dominated flow. A hole-dominated one 
is obtained by replacing $e\to-e$ and $\eta_H\to -\eta_H$ in Eqs.~\eqref{ODE2} and \eqref{ODE3} \cite{PhysRevLett.117.166601,Narozhny:2019mdg}. In this case, both the Lorentz and Hall viscous forces change signs but the relative sign is preserved, leading only to an overall  opposite Hall signal.

The transport coefficients $\sigma_Q$, $\eta$, and $\eta_H$ are known functions of $\mu/k_B T$ and $B_z$ \cite{Fritz:2008qct,Narozhny:2019mdg,Narozhny:2019ehg} (see also SM \cite{supMat}). 
Their most important features pertinent to our analysis are that $\eta$ and $\eta_H$ decrease for increasing $|B_z|$, while $\sigma_Q$ decreases with increasing $\mu/k_B T$.

Finally, we fix the boundary conditions of ${\bm v}$ by requiring a vanishing current outflow in the $y$-direction and a finite current along the channel boundaries. That is, 
\begin{equation}
\label{Eq:Vbcs}
v_x(\pm W/2) \pm l_s v_x'(\pm W/2)=  v_y(\pm W/2)=0~,
\end{equation} 
with the slip length $l_s$  [cf.~Fig.~\ref{fig:Cartoon}] parameterizing the diffusivity of the channel \cite{Kiselev:2018slipbdy}. 
The boundary conditions on $\delta\mu$ and $\delta T$ are fixed through the conservation of the total charge and energy within the channel, which implies
$\int_{-W/2}^{W/2} dy~ \delta n = 0= \int_{-W/2}^{W/2} dy~ \delta n_E$.

Inserting Eq.~\eqref{Eq:Vbcs} in Eqs.~\eqref{ODE1} and \eqref{ODE4}, one finds $J_y = P_y = v_y = 0$. With this, the velocity profile reads
\begin{equation}\label{Eq:Solution}
    v_x = -\frac{e n E_1l_G^2 }{\eta}\left(1-\frac{l_G^{\rm eff} \cosh \frac{y}{l_G}}{l_G \sinh \frac{W}{2l_G}}\right) ,
\end{equation}
while $\delta\mu$ and $\delta T$ are presented in the SM \cite{supMat}. We have defined the Gurzhi length $l_G = v_F \sqrt{\eta\tau_\text{MR}/w}$   \cite{Gurzhi:1963minimum,Gurzhi:1968hydrodynamic,Lucas:2018JPCM,PhysRevB.98.195143} and the effective Gurzhi length $l_G^\text{eff} = l_G[ \coth (W/2l_G)+l_s / l_G ]^{-1}$, which takes into account the effect of the channel geometry and the slip length on the flow.  As sketched in Fig.~\ref{fig:Cartoon}, $l_G^{\rm eff}$ reflects the effective segmentation of the channel of width $W$ into two boundary segments of width $l_G^\text{eff}$ and an interior segment of width $W-2l_G^\text{eff}$. For our range of parameters $l_G > 0.5 {\rm \mu m}$ at $B_z = 0$ and decreases as $B_z$ increases.

Some comments on $v_x$ are in order: At $B_z = 0$, $l_{ee} < W < l_G$, the system is in the Poiseuille regime and $v_x$ exhibits a parabolic dependence on $y$. Due to the large curvature of the flow, the Hall viscous effect becomes more important in this regime, as seen from Eq.~\eqref{ODE3}. 
When we increase the magnitude of $B_z$, the Gurzhi length $l_G$ decreases and drives the system into the Porous region $l_{ee} < l_G < W$ \cite{Sulpizio:2019NaturePoiseuille}. 
The velocity $v_x$ then exhibits a plateau with a maximum $v_{x,\rm max} = -enE_1v_F^2\tau_\text{MR}/w$ in the interior of the channel \cite{supMat}, rendering the Hall viscous effect negligible. 

We now draw our attention to the generalized Ohm's law for the conductivity matrix. Substituting Eq.~\eqref{Eq:Solution} into Eq.~\eqref{Currents} we obtain the average charge and heat currents along the channel. Imposing $J_y=P_y=v_y=0$, the Onsager relations along the channel reduce to  
\begin{align}\label{OnsagerRelation}
	&\begin{pmatrix}
			J^\text{avg}_x \\ Q^\text{avg}_x
	\end{pmatrix}
	=
	\int \frac{dy}{W}
	\begin{pmatrix}
		J_x \\
		Q_x
	\end{pmatrix}
	=
	\begin{pmatrix}
			\sigma & \alpha \\
			\bar\alpha T & \bar\kappa
	\end{pmatrix}
	\begin{pmatrix}
			E_x \\ - \nabla_x T
	\end{pmatrix} , \\
&
	\begin{pmatrix}
	\sigma & \alpha \\
	\bar\alpha T & \bar\kappa
	\end{pmatrix}
	=
	\begin{pmatrix}
	\sigma _Q+ \frac{e^2 n^2 v_F^2 }{w}\tau_\text{MR}^\text{avg} & \frac{\mu \sigma _Q}{e T}-\frac{e n s v_F^2 }{w}\tau_\text{MR}^\text{avg} \\
	\frac{\mu \sigma _Q}{e}-\frac{e n s T v_F^2 }{w}\tau_\text{MR}^\text{avg} & \frac{\mu^2 \sigma _Q}{e^2 T}+\frac{s^2 T v_F^2 }{w}\tau_\text{MR}^\text{avg} \\
\end{pmatrix} ,\nn
\end{align}
with the heat current $Q_x=P_xv_F-\frac{\mu}{-e} J_x$, electrical conductivity $\sigma$, thermoelectric conductivities $\alpha$ and $\bar\alpha$, and thermal conductivity $\bar\kappa$.  We have also defined the average momentum relaxation time $\tau_\text{MR}^\text{avg}=\tau_\text{MR} (1-2l_G^\text{eff}/W)$. It showcases the effect of finite boundaries on momentum relaxation. We note that $0< \tau_{\rm MR}^{\rm avg}\leq \tau_{\rm MR}$ holds for finite $W,\, l_G,\,l_s$.
In the SM \cite{supMat}, we show that Eq.~\eqref{OnsagerRelation} agrees with the Onsager relation at $W\to\infty$ limit \cite{Hartnoll:2007Nernst,Muller:PRBcyclotron}.

\textit{Hall voltage and hydrodynamic inverse Nernst effect.}\textemdash
Given the solution for $\delta\mu$, and $\delta T$, we can calculate explicity both the total Hall voltage $\Delta V$ and temperature gradient $\Delta T$ across the channel. Furthermore, the linear response assumption allows us to split both into two contributions; one due to the Lorentz force $\Delta V_B,\Delta T_B$ and another due to $\eta_H$ and the Hall viscous force $\Delta V_{\eta_H},\Delta T_{\eta_H}$. Namely, 
\begin{align}\label{HallVoltage}
	\Delta V &= (\delta \mu(W/2) -\delta \mu(-W/2))/(-e)=\Delta V_B+\Delta V_{\eta_H}, \nn\\
	\Delta V_B &= -\frac{W B_z }{ e w}\left(e^2  n v_F^2 \tau_\text{MR}^\text{avg}E_1+ \mu \sigma _QE_2\right), \\
	\Delta V_{\eta_H} &= -\frac{2  \mu n \eta _H l_G^\text{eff} }{\eta  w}E_1, \nn
\end{align}
and
\begin{align}\label{InverseNernst}
	\Delta T &= \delta T(W/2) -\delta T(-W/2)=\Delta T_{B}+\Delta T_{\eta_H} ,\\
	\Delta T_B &= \frac{ W B_z T  \sigma _Q}{w} E_2,
	\quad 
	\Delta T_{\eta_H} = \frac{2 e  n T\eta _H l_G^\text{eff} }{\eta w}E_1, 
	\nn
\end{align}
The decomposition in Eq.~\eqref{HallVoltage} showcases that the Hall viscous voltage, $\Delta V_{\eta_H}$, is generated from the curvature of the flow within the two boundary segments of size $l_G^{\rm eff}$. In contrast,
$\Delta V_B=-WB_zJ_x^\text{avg}$ takes the traditional form expected from the Hall effect. 
In our case, the current is given by Eq.~\eqref{OnsagerRelation} and contains two contributions; one from the quantum critical conductivity $\sigma_Q$ and another from a Drude conductivity, proportional to $\tau_{\rm MR}^{\rm avg}$, renormalized by the finite channel width and shear viscosity.
Remarkably, $\Delta T_B$ does not receive a contribution from momentum relaxation, due to the vanishing momentum density in $y$ direction [cf.~Eq.~\eqref{OnsagerRelation}]. 

\begin{figure}
	\includegraphics[height = 0.25\textwidth]{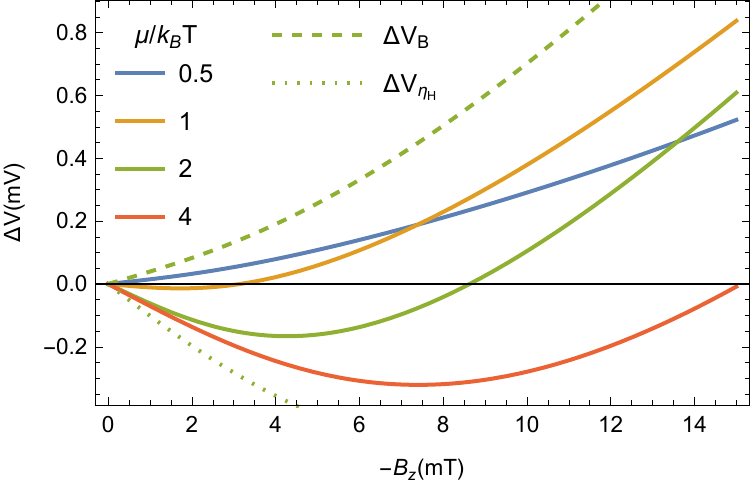}
	\includegraphics[height = 0.25\textwidth]{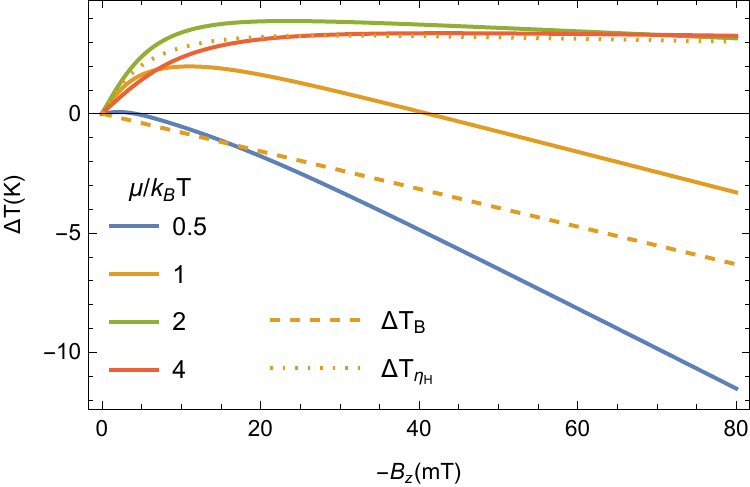}
	\caption{Hall voltages $\Delta V$ and temperature gradient $\Delta T$ as functions of the magnetic field $B_z$. 
	The other parameters are $T = 120~{\rm K}$, $E_x = -1000~{\rm V/m}$, $\nabla_x T = 0$, $W = 2~{\rm\mu m}$. 
	The two contributions $\Delta V_B$ and $\Delta V_{\eta_H}$ for $\mu = 2 k_B T$ and the two contributions $\Delta T_B$ and $\Delta T_{\eta_H}$ for $\mu = k_B T$ are shown (dashed and dotted lines).}
	\label{fig:HVB}
	\label{fig:HTB}
\end{figure}

Most notably, for electrons $\sgn(\eta_H)=-\sgn(B_z)$ results in $\Delta V_B$ ($\Delta T_B$) and $\Delta V_{\eta_H}$ ($\Delta T_{\eta_H}$) having opposite sign. Thus, the total Hall voltage or temperature gradient can become zero if the two contributions become equal. We can see from their ratio at $\nabla_x T=0$,
\begin{align}\label{HallRatio}
	\frac{\Delta V_{\eta_H}}{\Delta V_B}
	=\frac{2 e \mu  n \eta _H l_G^\text{eff}}{\eta  W B_z \left(e^2 n v_F^2 \tau_\text{MR}^\text{avg}+\mu  \sigma _Q\right)},\,
	\frac{\Delta T_{\eta_H}}{\Delta T_B}
	=\frac{2 e n \eta _H l_G^\text{eff}}{\eta W B_z \sigma_Q},
\end{align} 
that both the ratios can be comparable to unity only for small $l_s$, $W$, and $B_z$, which restricts ourselves to $l_s\ll W\sim 1{\rm \mu m}$. 
For increasing $l_s$, $W$, and $B_z$, $v_x$ becomes flatter and as a result the $\eta_H$ induced force becomes smaller.
Hence, while $\Delta V_{\eta_H}$ and $\Delta T_{\eta_H}$ saturate according to Eqs.~\eqref{HallVoltage} and \eqref{InverseNernst}, $|\Delta V_{B}|$ and $|\Delta T_{B}|$ keep growing with increasing $|B_z|$ and, eventually, $\Delta V$ and $\Delta T$ are dominated by $\Delta V_{B}$ and $\Delta T_{B}$, respectively. This leads to the \textit{non-monotonic} behavior in Fig.~\ref{fig:HVB} 
and in particular to a zero of both $\Delta V$ and $\Delta T$ at a non-vanishing value of $B_z$.
We note that the non-monotonic behavior appears only in the regime $\mu \gtrsim k_B T$, but is absent or very weak for $\mu < k_B T$. 
This can be seen from the limit $\mu/k_BT \ll 1$ in which $n\to 0$ and both ratios in Eq.~\eqref{HallRatio} become equal and less than $1$ within our parameters range.

It is important to emphasize, however, that the non-monotonic behavior of $\Delta V$ is different to the one of $\Delta T$, due to the Drude contribution in $\Delta V_B$. First, the critical magnetic field for $\Delta V$ is smaller than that for $\Delta T$. Second, since $\sigma_Q$ is small in the Fermi liquid regime $\mu \gg k_BT$, the ratio $\Delta T_{\eta_H}/\Delta T_B$ is enhanced and the Hall viscous effect dominates the inverse Nernst effect, which is different for $\Delta V_{\eta_H}/\Delta V_B$  \cite{supMat}.

\begin{figure}
	\includegraphics[width = 0.4\textwidth]{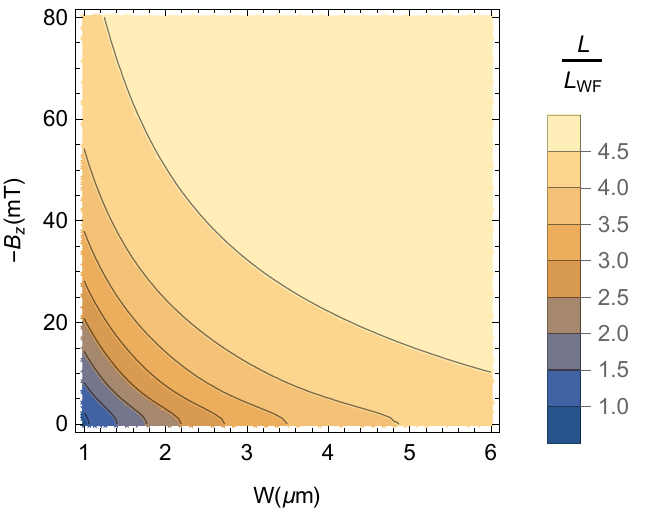}
	\caption{The Lorenz ratio $L$ in units of $L_{\rm WF}$ as a function of the width $W$ and the magnetic field $B_z$ at $\mu = 0.5~k_B T$ and $T=120K$. 
	When $W\to \infty$, it approaches $L_\infty / L_{\rm WF} \approx 4.9$. 
	} 
	\label{fig:lorenz}
\end{figure}

\textit{Suppression of the Lorenz ratio.}\textemdash
Our simulations predict a violation of the Wiedemann-Franz law in the Dirac regime which increases with increasing magnetic field and channel width.
The Lorenz ratio is defined as
\begin{align}\label{LorentzRatio}
	L &= \frac{\kappa}{\sigma T}=\frac{L_0}{(1+(n/n_0)^2)^2},
	\\
	L_0 &= \frac{w v_F^2 \tau_\text{MR}^\text{avg}}{T^2 \sigma _Q},
	\quad
	n_0^2=\frac{w \sigma _Q }{e^2 v_F^2\tau_\text{MR}^\text{avg}} , \nn
\end{align}
with the thermal conductivity $\kappa$ defined by $Q^\text{avg}_x = -\kappa \nabla_x T$ at $J^\text{avg}_x=0$ characterizing the pure heat flow.
The Wiedemann-Franz law $L_{\rm WF} = \pi^2 k_B^2 / (3e^2)$ \cite{Franz:1853WFLaw}, which is valid for non-interacting systems, was found to be violated for strongly correlated systems \cite{LorentzRatioNFL,Crossno:LorentzRatio}. 
In our channel setup, both the Onsager relation and the Lorenz ratio $L$ take their hydrodynamic form in boundaryless graphene \cite{Crossno:LorentzRatio}, with the momentum relaxation time $\tau_\text{MR}$ replaced by the average $\tau_\text{MR}^\text{avg}$. 
As shown in Fig.~\ref{fig:lorenz}, the Lorenz ratio $L$ implicitly depends on $B_z$ and $W$ via $\tau_\text{MR}^\text{avg}$.
If we increase $W$ or $B_z$ such that $\tau_\text{MR}^\text{avg}\to\tau_\text{MR}$, $L$ also approaches its value $L_\infty$ for $W\to\infty$.

\textit{Conclusion.}\textemdash
By considering a channel sample of graphene subjected to electric and magnetic fields, we have shown that the electronic fluid is characterized by two effective scales due to the finite channel width, i.e., the effective Gurzhi length $l_G^\text{eff}$ and average momentum relaxation time $\tau_\text{MR}^\text{eff}$.
The generated Hall voltage in Eq.~\eqref{HallVoltage} and temperature gradient in Eq.~\eqref{InverseNernst} exhibit a non-monotonic dependence on the magnetic field that reflects the competition between the Hall viscous and Lorentz effects, where the former is dominant at small and the latter at large magnetic field. 
Furthermore, one can directly relate the hydrodynamic inverse Nernst signal in the Fermi liquid regime to its Hall viscous contribution since the Lorentz contribution is suppressed by the small $\sigma_Q$.
Finally, we find that the Lorenz ratio in Eq.~\eqref{LorentzRatio} is suppressed by a finite width $W$ through $\tau_\text{MR}^\text{eff}$, while a magnetic field leads to an enhancement of Lorenz ratio.
Possible extensions of this work include calculating the Hall signal via kinetic theory \cite{Scaffidi:PRL2017}, investigating the effect of out-of-plane magnetic fields and the Hall viscosity on the preturbulent vortex shedding in graphene  \cite{mendoza2011preturbulent} as well as on fully developed turbulence in Kagome metals \cite{DiSante:2019zrd}, analyzing the interplay of parity-breaking Hall viscosity with the parity anomaly present, e.g., in quantum anomalous Hall systems \cite{bottcher2019survival,tutschku2020momentum,Tutschku:2020rjq}, an analysis of the AC version of the Poiseulle flow \cite{moessner2018pulsating} including the Hall viscosity, as well as investigating the inverse Nernst effect in multiterminal \cite{Geim:2015PRBNonlocalHydro,Geim:2016Negative} or Corbino geometries \cite{holder2019unified}. 
Finally, it was recently suggested that collective effects due to plasmons may also modify thermoelectric transport in graphene \cite{https://doi.org/10.48550/arxiv.2206.09687,2022arXiv220609694P}. 
It is of interest to examine if including plasmon dynamics to the hydrodynamic equations, in particular including a magnetic field, lead to discernible differences in the Hall voltage and inverse Nernst effect. 

\begin{acknowledgments}
We thank Amir Yacoby and his group, in particular Ziwei Qiu, for useful discussions. We gratefully acknowledge support from the DFG via SFB 1170 "Topological and Correlated Electronics at Surfaces and Interfaces" (project id 258499086), and via the W{\"u}rzburg-Dresden Cluster of Excellence on Complexity and Topology in Quantum Matter — ct.qmat (EXC 2147, project-id 390858490). In addition, D.R.F.~has been partially supported by the Netherlands Organization for Scientific Research/Ministry of Science and Education (NWO/OCW), while I.M.~has also been partially supported by the ``Curiosity Driven Grant 2020'' of the University of Genoa and by the INFN Scientific Initiatives SFT: ``Statistical Field Theory, Low-Dimensional Systems, Integrable Models and Applications''.
\end{acknowledgments}

\bibliographystyle{apsrev}
\bibliography{ref}

\begin{thebibliography}{79}
\expandafter\ifx\csname natexlab\endcsname\relax\def\natexlab#1{#1}\fi
\expandafter\ifx\csname bibnamefont\endcsname\relax
  \def\bibnamefont#1{#1}\fi
\expandafter\ifx\csname bibfnamefont\endcsname\relax
  \def\bibfnamefont#1{#1}\fi
\expandafter\ifx\csname citenamefont\endcsname\relax
  \def\citenamefont#1{#1}\fi
\expandafter\ifx\csname url\endcsname\relax
  \def\url#1{\texttt{#1}}\fi
\expandafter\ifx\csname urlprefix\endcsname\relax\def\urlprefix{URL }\fi
\providecommand{\bibinfo}[2]{#2}
\providecommand{\eprint}[2][]{\url{#2}}

\bibitem[{\citenamefont{Moll et~al.}(2016)\citenamefont{Moll, Kushwaha, Nandi,
  Schmidt, and Mackenzie}}]{science:aac8385}
\bibinfo{author}{\bibfnamefont{P.~J.~W.} \bibnamefont{Moll}},
  \bibinfo{author}{\bibfnamefont{P.}~\bibnamefont{Kushwaha}},
  \bibinfo{author}{\bibfnamefont{N.}~\bibnamefont{Nandi}},
  \bibinfo{author}{\bibfnamefont{B.}~\bibnamefont{Schmidt}}, \bibnamefont{and}
  \bibinfo{author}{\bibfnamefont{A.~P.} \bibnamefont{Mackenzie}},
  \bibinfo{journal}{Science} \textbf{\bibinfo{volume}{351}},
  \bibinfo{pages}{1061} (\bibinfo{year}{2016}).

\bibitem[{\citenamefont{Crossno et~al.}(2016)\citenamefont{Crossno, Shi, Wang,
  Liu, Harzheim, Lucas, Sachdev, Kim, Taniguchi, Watanabe
  et~al.}}]{Crossno:LorentzRatio}
\bibinfo{author}{\bibfnamefont{J.}~\bibnamefont{Crossno}},
  \bibinfo{author}{\bibfnamefont{J.~K.} \bibnamefont{Shi}},
  \bibinfo{author}{\bibfnamefont{K.}~\bibnamefont{Wang}},
  \bibinfo{author}{\bibfnamefont{X.}~\bibnamefont{Liu}},
  \bibinfo{author}{\bibfnamefont{A.}~\bibnamefont{Harzheim}},
  \bibinfo{author}{\bibfnamefont{A.}~\bibnamefont{Lucas}},
  \bibinfo{author}{\bibfnamefont{S.}~\bibnamefont{Sachdev}},
  \bibinfo{author}{\bibfnamefont{P.}~\bibnamefont{Kim}},
  \bibinfo{author}{\bibfnamefont{T.}~\bibnamefont{Taniguchi}},
  \bibinfo{author}{\bibfnamefont{K.}~\bibnamefont{Watanabe}},
  \bibnamefont{et~al.}, \bibinfo{journal}{Science}
  \textbf{\bibinfo{volume}{351}}, \bibinfo{pages}{1058} (\bibinfo{year}{2016}).

\bibitem[{\citenamefont{{Sulpizio} et~al.}(2019)\citenamefont{{Sulpizio},
  {Ella}, {Rozen}, {Birkbeck}, {Perello}, {Dutta}, {Ben-Shalom}, {Taniguchi},
  {Watanabe}, {Holder} et~al.}}]{Sulpizio:2019NaturePoiseuille}
\bibinfo{author}{\bibfnamefont{J.~A.} \bibnamefont{{Sulpizio}}},
  \bibinfo{author}{\bibfnamefont{L.}~\bibnamefont{{Ella}}},
  \bibinfo{author}{\bibfnamefont{A.}~\bibnamefont{{Rozen}}},
  \bibinfo{author}{\bibfnamefont{J.}~\bibnamefont{{Birkbeck}}},
  \bibinfo{author}{\bibfnamefont{D.~J.} \bibnamefont{{Perello}}},
  \bibinfo{author}{\bibfnamefont{D.}~\bibnamefont{{Dutta}}},
  \bibinfo{author}{\bibfnamefont{M.}~\bibnamefont{{Ben-Shalom}}},
  \bibinfo{author}{\bibfnamefont{T.}~\bibnamefont{{Taniguchi}}},
  \bibinfo{author}{\bibfnamefont{K.}~\bibnamefont{{Watanabe}}},
  \bibinfo{author}{\bibfnamefont{T.}~\bibnamefont{{Holder}}},
  \bibnamefont{et~al.}, \bibinfo{journal}{\nat} \textbf{\bibinfo{volume}{576}},
  \bibinfo{pages}{75} (\bibinfo{year}{2019}).

\bibitem[{\citenamefont{{Ku} et~al.}(2020)\citenamefont{{Ku}, {Zhou}, {Li},
  {Shin}, {Shi}, {Burch}, {Anderson}, {Pierce}, {Xie}, {Hamo}
  et~al.}}]{Ku:2020NatureViscousFlow}
\bibinfo{author}{\bibfnamefont{M.~J.~H.} \bibnamefont{{Ku}}},
  \bibinfo{author}{\bibfnamefont{T.~X.} \bibnamefont{{Zhou}}},
  \bibinfo{author}{\bibfnamefont{Q.}~\bibnamefont{{Li}}},
  \bibinfo{author}{\bibfnamefont{Y.~J.} \bibnamefont{{Shin}}},
  \bibinfo{author}{\bibfnamefont{J.~K.} \bibnamefont{{Shi}}},
  \bibinfo{author}{\bibfnamefont{C.}~\bibnamefont{{Burch}}},
  \bibinfo{author}{\bibfnamefont{L.~E.} \bibnamefont{{Anderson}}},
  \bibinfo{author}{\bibfnamefont{A.~T.} \bibnamefont{{Pierce}}},
  \bibinfo{author}{\bibfnamefont{Y.}~\bibnamefont{{Xie}}},
  \bibinfo{author}{\bibfnamefont{A.}~\bibnamefont{{Hamo}}},
  \bibnamefont{et~al.}, \bibinfo{journal}{\nat} \textbf{\bibinfo{volume}{583}},
  \bibinfo{pages}{537} (\bibinfo{year}{2020}).

\bibitem[{\citenamefont{Jenkins et~al.}(2020)\citenamefont{Jenkins, Baumann,
  Zhou, Meynell, Yang, Watanabe, Taniguchi, Lucas, Young, and
  Jayich}}]{Jenkins:2020Imaging}
\bibinfo{author}{\bibfnamefont{A.}~\bibnamefont{Jenkins}},
  \bibinfo{author}{\bibfnamefont{S.}~\bibnamefont{Baumann}},
  \bibinfo{author}{\bibfnamefont{H.}~\bibnamefont{Zhou}},
  \bibinfo{author}{\bibfnamefont{S.~A.} \bibnamefont{Meynell}},
  \bibinfo{author}{\bibfnamefont{D.}~\bibnamefont{Yang}},
  \bibinfo{author}{\bibfnamefont{K.}~\bibnamefont{Watanabe}},
  \bibinfo{author}{\bibfnamefont{T.}~\bibnamefont{Taniguchi}},
  \bibinfo{author}{\bibfnamefont{A.}~\bibnamefont{Lucas}},
  \bibinfo{author}{\bibfnamefont{A.~F.} \bibnamefont{Young}}, \bibnamefont{and}
  \bibinfo{author}{\bibfnamefont{A.~C.~B.} \bibnamefont{Jayich}},
  \bibinfo{journal}{arXiv:2002.05065}  (\bibinfo{year}{2020}).

\bibitem[{\citenamefont{Vool et~al.}(2021)\citenamefont{Vool, Hamo, Varnavides,
  Wang, Zhou, Kumar, Dovzhenko, Qiu, Garcia, Pierce et~al.}}]{Vool2021}
\bibinfo{author}{\bibfnamefont{U.}~\bibnamefont{Vool}},
  \bibinfo{author}{\bibfnamefont{A.}~\bibnamefont{Hamo}},
  \bibinfo{author}{\bibfnamefont{G.}~\bibnamefont{Varnavides}},
  \bibinfo{author}{\bibfnamefont{Y.}~\bibnamefont{Wang}},
  \bibinfo{author}{\bibfnamefont{T.~X.} \bibnamefont{Zhou}},
  \bibinfo{author}{\bibfnamefont{N.}~\bibnamefont{Kumar}},
  \bibinfo{author}{\bibfnamefont{Y.}~\bibnamefont{Dovzhenko}},
  \bibinfo{author}{\bibfnamefont{Z.}~\bibnamefont{Qiu}},
  \bibinfo{author}{\bibfnamefont{C.~A.~C.} \bibnamefont{Garcia}},
  \bibinfo{author}{\bibfnamefont{A.~T.} \bibnamefont{Pierce}},
  \bibnamefont{et~al.}, \bibinfo{journal}{Nat. Phys.}
  \textbf{\bibinfo{volume}{17}}, \bibinfo{pages}{1216} (\bibinfo{year}{2021}).

\bibitem[{\citenamefont{Bandurin et~al.}(2016)\citenamefont{Bandurin, Torre,
  Kumar, Ben~Shalom, Tomadin, Principi, Auton, Khestanova, Novoselov,
  Grigorieva et~al.}}]{Geim:2016Negative}
\bibinfo{author}{\bibfnamefont{D.}~\bibnamefont{Bandurin}},
  \bibinfo{author}{\bibfnamefont{I.}~\bibnamefont{Torre}},
  \bibinfo{author}{\bibfnamefont{R.~K.} \bibnamefont{Kumar}},
  \bibinfo{author}{\bibfnamefont{M.}~\bibnamefont{Ben~Shalom}},
  \bibinfo{author}{\bibfnamefont{A.}~\bibnamefont{Tomadin}},
  \bibinfo{author}{\bibfnamefont{A.}~\bibnamefont{Principi}},
  \bibinfo{author}{\bibfnamefont{G.}~\bibnamefont{Auton}},
  \bibinfo{author}{\bibfnamefont{E.}~\bibnamefont{Khestanova}},
  \bibinfo{author}{\bibfnamefont{K.}~\bibnamefont{Novoselov}},
  \bibinfo{author}{\bibfnamefont{I.}~\bibnamefont{Grigorieva}},
  \bibnamefont{et~al.}, \bibinfo{journal}{Science}
  \textbf{\bibinfo{volume}{351}}, \bibinfo{pages}{1055} (\bibinfo{year}{2016}).

\bibitem[{\citenamefont{Gallagher et~al.}(2019)\citenamefont{Gallagher, Yang,
  Lyu, Tian, Kou, Zhang, Watanabe, Taniguchi, and
  Wang}}]{Gallagher:2019MontumRelaxation}
\bibinfo{author}{\bibfnamefont{P.}~\bibnamefont{Gallagher}},
  \bibinfo{author}{\bibfnamefont{C.-S.} \bibnamefont{Yang}},
  \bibinfo{author}{\bibfnamefont{T.}~\bibnamefont{Lyu}},
  \bibinfo{author}{\bibfnamefont{F.}~\bibnamefont{Tian}},
  \bibinfo{author}{\bibfnamefont{R.}~\bibnamefont{Kou}},
  \bibinfo{author}{\bibfnamefont{H.}~\bibnamefont{Zhang}},
  \bibinfo{author}{\bibfnamefont{K.}~\bibnamefont{Watanabe}},
  \bibinfo{author}{\bibfnamefont{T.}~\bibnamefont{Taniguchi}},
  \bibnamefont{and} \bibinfo{author}{\bibfnamefont{F.}~\bibnamefont{Wang}},
  \bibinfo{journal}{Science} \textbf{\bibinfo{volume}{364}},
  \bibinfo{pages}{158} (\bibinfo{year}{2019}).

\bibitem[{\citenamefont{{Berdyugin} et~al.}(2019)\citenamefont{{Berdyugin},
  {Xu}, {Pellegrino}, {Krishna Kumar}, {Principi}, {Torre}, {Ben Shalom},
  {Taniguchi}, {Watanabe}, {Grigorieva} et~al.}}]{Berdyugin:2019SciHall}
\bibinfo{author}{\bibfnamefont{A.~I.} \bibnamefont{{Berdyugin}}},
  \bibinfo{author}{\bibfnamefont{S.~G.} \bibnamefont{{Xu}}},
  \bibinfo{author}{\bibfnamefont{F.~M.~D.} \bibnamefont{{Pellegrino}}},
  \bibinfo{author}{\bibfnamefont{R.}~\bibnamefont{{Krishna Kumar}}},
  \bibinfo{author}{\bibfnamefont{A.}~\bibnamefont{{Principi}}},
  \bibinfo{author}{\bibfnamefont{I.}~\bibnamefont{{Torre}}},
  \bibinfo{author}{\bibfnamefont{M.}~\bibnamefont{{Ben Shalom}}},
  \bibinfo{author}{\bibfnamefont{T.}~\bibnamefont{{Taniguchi}}},
  \bibinfo{author}{\bibfnamefont{K.}~\bibnamefont{{Watanabe}}},
  \bibinfo{author}{\bibfnamefont{I.~V.} \bibnamefont{{Grigorieva}}},
  \bibnamefont{et~al.}, \bibinfo{journal}{Science}
  \textbf{\bibinfo{volume}{364}}, \bibinfo{pages}{162} (\bibinfo{year}{2019}).

\bibitem[{\citenamefont{Ella et~al.}(2019)\citenamefont{Ella, Rozen, Birkbeck,
  Ben-Shalom, Perello, Zultak, Taniguchi, Watanabe, Geim, Ilani
  et~al.}}]{ella2019simultaneous}
\bibinfo{author}{\bibfnamefont{L.}~\bibnamefont{Ella}},
  \bibinfo{author}{\bibfnamefont{A.}~\bibnamefont{Rozen}},
  \bibinfo{author}{\bibfnamefont{J.}~\bibnamefont{Birkbeck}},
  \bibinfo{author}{\bibfnamefont{M.}~\bibnamefont{Ben-Shalom}},
  \bibinfo{author}{\bibfnamefont{D.}~\bibnamefont{Perello}},
  \bibinfo{author}{\bibfnamefont{J.}~\bibnamefont{Zultak}},
  \bibinfo{author}{\bibfnamefont{T.}~\bibnamefont{Taniguchi}},
  \bibinfo{author}{\bibfnamefont{K.}~\bibnamefont{Watanabe}},
  \bibinfo{author}{\bibfnamefont{A.~K.} \bibnamefont{Geim}},
  \bibinfo{author}{\bibfnamefont{S.}~\bibnamefont{Ilani}},
  \bibnamefont{et~al.}, \bibinfo{journal}{Nat. Nanotechnol.}
  \textbf{\bibinfo{volume}{14}}, \bibinfo{pages}{480} (\bibinfo{year}{2019}).

\bibitem[{\citenamefont{Bandurin et~al.}(2018)\citenamefont{Bandurin, Shytov,
  Levitov, Kumar, Berdyugin, Ben~Shalom, Grigorieva, Geim, and
  Falkovich}}]{bandurin2018fluidity}
\bibinfo{author}{\bibfnamefont{D.~A.} \bibnamefont{Bandurin}},
  \bibinfo{author}{\bibfnamefont{A.~V.} \bibnamefont{Shytov}},
  \bibinfo{author}{\bibfnamefont{L.~S.} \bibnamefont{Levitov}},
  \bibinfo{author}{\bibfnamefont{R.~K.} \bibnamefont{Kumar}},
  \bibinfo{author}{\bibfnamefont{A.~I.} \bibnamefont{Berdyugin}},
  \bibinfo{author}{\bibfnamefont{M.}~\bibnamefont{Ben~Shalom}},
  \bibinfo{author}{\bibfnamefont{I.~V.} \bibnamefont{Grigorieva}},
  \bibinfo{author}{\bibfnamefont{A.~K.} \bibnamefont{Geim}}, \bibnamefont{and}
  \bibinfo{author}{\bibfnamefont{G.}~\bibnamefont{Falkovich}},
  \bibinfo{journal}{Nat. Commun.} \textbf{\bibinfo{volume}{9}},
  \bibinfo{pages}{1} (\bibinfo{year}{2018}).

\bibitem[{\citenamefont{{Lucas} and {Chung Fong}}(2018)}]{Lucas:2018JPCM}
\bibinfo{author}{\bibfnamefont{A.}~\bibnamefont{{Lucas}}} \bibnamefont{and}
  \bibinfo{author}{\bibfnamefont{K.}~\bibnamefont{{Chung Fong}}},
  \bibinfo{journal}{J. Phys.: Condens. Matter} \textbf{\bibinfo{volume}{30}},
  \bibinfo{eid}{053001} (\bibinfo{year}{2018}).

\bibitem[{\citenamefont{{Narozhny} et~al.}(2017)\citenamefont{{Narozhny},
  {Gornyi}, {Mirlin}, and {Schmalian}}}]{Narozhny:2017HydroGraphene}
\bibinfo{author}{\bibfnamefont{B.~N.} \bibnamefont{{Narozhny}}},
  \bibinfo{author}{\bibfnamefont{I.~V.} \bibnamefont{{Gornyi}}},
  \bibinfo{author}{\bibfnamefont{A.~D.} \bibnamefont{{Mirlin}}},
  \bibnamefont{and}
  \bibinfo{author}{\bibfnamefont{J.}~\bibnamefont{{Schmalian}}},
  \bibinfo{journal}{Ann. Phys.} \textbf{\bibinfo{volume}{529}},
  \bibinfo{pages}{1700043} (\bibinfo{year}{2017}).

\bibitem[{\citenamefont{Narozhny}(2022)}]{Narozhny2022}
\bibinfo{author}{\bibfnamefont{B.~N.} \bibnamefont{Narozhny}},
  \bibinfo{journal}{Riv. Nuovo Cim.}  (\bibinfo{year}{2022}).

\bibitem[{\citenamefont{Torre et~al.}(2015)\citenamefont{Torre, Tomadin, Geim,
  and Polini}}]{Geim:2015PRBNonlocalHydro}
\bibinfo{author}{\bibfnamefont{I.}~\bibnamefont{Torre}},
  \bibinfo{author}{\bibfnamefont{A.}~\bibnamefont{Tomadin}},
  \bibinfo{author}{\bibfnamefont{A.~K.} \bibnamefont{Geim}}, \bibnamefont{and}
  \bibinfo{author}{\bibfnamefont{M.}~\bibnamefont{Polini}},
  \bibinfo{journal}{Phys. Rev. B} \textbf{\bibinfo{volume}{92}},
  \bibinfo{pages}{165433} (\bibinfo{year}{2015}).

\bibitem[{\citenamefont{Molenkamp and de~Jong}(1994)}]{Molenkamp:PRBGaAs}
\bibinfo{author}{\bibfnamefont{L.~W.} \bibnamefont{Molenkamp}}
  \bibnamefont{and} \bibinfo{author}{\bibfnamefont{M.~J.~M.}
  \bibnamefont{de~Jong}}, \bibinfo{journal}{Phys. Rev. B}
  \textbf{\bibinfo{volume}{49}}, \bibinfo{pages}{5038} (\bibinfo{year}{1994}).

\bibitem[{\citenamefont{de~Jong and Molenkamp}(1995)}]{Jong:PRBGaAs}
\bibinfo{author}{\bibfnamefont{M.~J.~M.} \bibnamefont{de~Jong}}
  \bibnamefont{and} \bibinfo{author}{\bibfnamefont{L.~W.}
  \bibnamefont{Molenkamp}}, \bibinfo{journal}{Phys. Rev. B}
  \textbf{\bibinfo{volume}{51}}, \bibinfo{pages}{13389} (\bibinfo{year}{1995}).

\bibitem[{\citenamefont{Gusev et~al.}(2018)\citenamefont{Gusev, Levin,
  Levinson, and Bakarov}}]{gusev2018viscous}
\bibinfo{author}{\bibfnamefont{G.}~\bibnamefont{Gusev}},
  \bibinfo{author}{\bibfnamefont{A.}~\bibnamefont{Levin}},
  \bibinfo{author}{\bibfnamefont{E.}~\bibnamefont{Levinson}}, \bibnamefont{and}
  \bibinfo{author}{\bibfnamefont{A.}~\bibnamefont{Bakarov}},
  \bibinfo{journal}{AIP Adv.} \textbf{\bibinfo{volume}{8}},
  \bibinfo{pages}{025318} (\bibinfo{year}{2018}).

\bibitem[{\citenamefont{Alekseev}(2016)}]{PhysRevLett.117.166601}
\bibinfo{author}{\bibfnamefont{P.~S.} \bibnamefont{Alekseev}},
  \bibinfo{journal}{Phys. Rev. Lett.} \textbf{\bibinfo{volume}{117}},
  \bibinfo{pages}{166601} (\bibinfo{year}{2016}).

\bibitem[{\citenamefont{Alekseev et~al.}(2015)\citenamefont{Alekseev, Dmitriev,
  Gornyi, Kachorovskii, Narozhny, Sch\"utt, and Titov}}]{Alekseev:2015PRL}
\bibinfo{author}{\bibfnamefont{P.~S.} \bibnamefont{Alekseev}},
  \bibinfo{author}{\bibfnamefont{A.~P.} \bibnamefont{Dmitriev}},
  \bibinfo{author}{\bibfnamefont{I.~V.} \bibnamefont{Gornyi}},
  \bibinfo{author}{\bibfnamefont{V.~Y.} \bibnamefont{Kachorovskii}},
  \bibinfo{author}{\bibfnamefont{B.~N.} \bibnamefont{Narozhny}},
  \bibinfo{author}{\bibfnamefont{M.}~\bibnamefont{Sch\"utt}}, \bibnamefont{and}
  \bibinfo{author}{\bibfnamefont{M.}~\bibnamefont{Titov}},
  \bibinfo{journal}{Phys. Rev. Lett.} \textbf{\bibinfo{volume}{114}},
  \bibinfo{pages}{156601} (\bibinfo{year}{2015}).

\bibitem[{\citenamefont{Alekseev et~al.}(2018)\citenamefont{Alekseev, Dmitriev,
  Gornyi, Kachorovskii, Narozhny, and Titov}}]{Alekseev:magnetoresistance}
\bibinfo{author}{\bibfnamefont{P.~S.} \bibnamefont{Alekseev}},
  \bibinfo{author}{\bibfnamefont{A.~P.} \bibnamefont{Dmitriev}},
  \bibinfo{author}{\bibfnamefont{I.~V.} \bibnamefont{Gornyi}},
  \bibinfo{author}{\bibfnamefont{V.~Y.} \bibnamefont{Kachorovskii}},
  \bibinfo{author}{\bibfnamefont{B.~N.} \bibnamefont{Narozhny}},
  \bibnamefont{and} \bibinfo{author}{\bibfnamefont{M.}~\bibnamefont{Titov}},
  \bibinfo{journal}{Phys. Rev. B} \textbf{\bibinfo{volume}{97}},
  \bibinfo{pages}{085109} (\bibinfo{year}{2018}).

\bibitem[{\citenamefont{Alekseev et~al.}(2017)\citenamefont{Alekseev, Dmitriev,
  Gornyi, Kachorovskii, Narozhny, Sch\"utt, and
  Titov}}]{Alekseev:2017PRBMagnetoresistance}
\bibinfo{author}{\bibfnamefont{P.~S.} \bibnamefont{Alekseev}},
  \bibinfo{author}{\bibfnamefont{A.~P.} \bibnamefont{Dmitriev}},
  \bibinfo{author}{\bibfnamefont{I.~V.} \bibnamefont{Gornyi}},
  \bibinfo{author}{\bibfnamefont{V.~Y.} \bibnamefont{Kachorovskii}},
  \bibinfo{author}{\bibfnamefont{B.~N.} \bibnamefont{Narozhny}},
  \bibinfo{author}{\bibfnamefont{M.}~\bibnamefont{Sch\"utt}}, \bibnamefont{and}
  \bibinfo{author}{\bibfnamefont{M.}~\bibnamefont{Titov}},
  \bibinfo{journal}{Phys. Rev. B} \textbf{\bibinfo{volume}{95}},
  \bibinfo{pages}{165410} (\bibinfo{year}{2017}).

\bibitem[{\citenamefont{Avron et~al.}(1995)\citenamefont{Avron, Seiler, and
  Zograf}}]{Avron:1995PRLHall}
\bibinfo{author}{\bibfnamefont{J.~E.} \bibnamefont{Avron}},
  \bibinfo{author}{\bibfnamefont{R.}~\bibnamefont{Seiler}}, \bibnamefont{and}
  \bibinfo{author}{\bibfnamefont{P.~G.} \bibnamefont{Zograf}},
  \bibinfo{journal}{Phys. Rev. Lett.} \textbf{\bibinfo{volume}{75}},
  \bibinfo{pages}{697} (\bibinfo{year}{1995}).

\bibitem[{\citenamefont{Jensen et~al.}(2012)\citenamefont{Jensen, Kaminski,
  Kovtun, Meyer, Ritz, and Yarom}}]{Jensen:2011xb}
\bibinfo{author}{\bibfnamefont{K.}~\bibnamefont{Jensen}},
  \bibinfo{author}{\bibfnamefont{M.}~\bibnamefont{Kaminski}},
  \bibinfo{author}{\bibfnamefont{P.}~\bibnamefont{Kovtun}},
  \bibinfo{author}{\bibfnamefont{R.}~\bibnamefont{Meyer}},
  \bibinfo{author}{\bibfnamefont{A.}~\bibnamefont{Ritz}}, \bibnamefont{and}
  \bibinfo{author}{\bibfnamefont{A.}~\bibnamefont{Yarom}}, \bibinfo{journal}{J.
  High Energ. Phys.} \textbf{\bibinfo{volume}{05}}, \bibinfo{pages}{102}
  (\bibinfo{year}{2012}).

\bibitem[{\citenamefont{{Hoyos}}(2014)}]{Hoyos:2014Hall}
\bibinfo{author}{\bibfnamefont{C.}~\bibnamefont{{Hoyos}}},
  \bibinfo{journal}{Int. J. Mod. Phys. B} \textbf{\bibinfo{volume}{28}},
  \bibinfo{eid}{1430007} (\bibinfo{year}{2014}).

\bibitem[{\citenamefont{{Matthaiakakis}
  et~al.}(2020)\citenamefont{{Matthaiakakis}, {Rodr{\'\i}guez Fern{\'a}ndez},
  {Tutschku}, {Hankiewicz}, {Erdmenger}, and {Meyer}}}]{Matthaiakakis:2020PRB}
\bibinfo{author}{\bibfnamefont{I.}~\bibnamefont{{Matthaiakakis}}},
  \bibinfo{author}{\bibfnamefont{D.}~\bibnamefont{{Rodr{\'\i}guez
  Fern{\'a}ndez}}},
  \bibinfo{author}{\bibfnamefont{C.}~\bibnamefont{{Tutschku}}},
  \bibinfo{author}{\bibfnamefont{E.~M.} \bibnamefont{{Hankiewicz}}},
  \bibinfo{author}{\bibfnamefont{J.}~\bibnamefont{{Erdmenger}}},
  \bibnamefont{and} \bibinfo{author}{\bibfnamefont{R.}~\bibnamefont{{Meyer}}},
  \bibinfo{journal}{\prb} \textbf{\bibinfo{volume}{101}}, \bibinfo{eid}{045423}
  (\bibinfo{year}{2020}).

\bibitem[{Foo({\natexlab{a}})}]{Footnote1}
\bibinfo{note}{{Note that we are prohibited from going to arbitrarily small
  densities since charge puddles are present in graphene
  \cite{Lucas:2018JPCM}.}}

\bibitem[{\citenamefont{{M{\"u}ller} et~al.}(2008)\citenamefont{{M{\"u}ller},
  {Fritz}, and {Sachdev}}}]{Markus:2008quantumcritical}
\bibinfo{author}{\bibfnamefont{M.}~\bibnamefont{{M{\"u}ller}}},
  \bibinfo{author}{\bibfnamefont{L.}~\bibnamefont{{Fritz}}}, \bibnamefont{and}
  \bibinfo{author}{\bibfnamefont{S.}~\bibnamefont{{Sachdev}}},
  \bibinfo{journal}{\prb} \textbf{\bibinfo{volume}{78}}, \bibinfo{eid}{115406}
  (\bibinfo{year}{2008}).

\bibitem[{\citenamefont{Behnia and Aubin}(2016)}]{Behnia_2016}
\bibinfo{author}{\bibfnamefont{K.}~\bibnamefont{Behnia}} \bibnamefont{and}
  \bibinfo{author}{\bibfnamefont{H.}~\bibnamefont{Aubin}},
  \bibinfo{journal}{Rep. Prog. Phys.} \textbf{\bibinfo{volume}{79}},
  \bibinfo{pages}{046502} (\bibinfo{year}{2016}).

\bibitem[{\citenamefont{Sakai et~al.}(2018)\citenamefont{Sakai, Mizuta,
  Nugroho, Sihombing, Koretsune, Suzuki, Takemori, Ishii, Nishio-Hamane, Arita
  et~al.}}]{Sakai_2018}
\bibinfo{author}{\bibfnamefont{A.}~\bibnamefont{Sakai}},
  \bibinfo{author}{\bibfnamefont{Y.~P.} \bibnamefont{Mizuta}},
  \bibinfo{author}{\bibfnamefont{A.~A.} \bibnamefont{Nugroho}},
  \bibinfo{author}{\bibfnamefont{R.}~\bibnamefont{Sihombing}},
  \bibinfo{author}{\bibfnamefont{T.}~\bibnamefont{Koretsune}},
  \bibinfo{author}{\bibfnamefont{M.-T.} \bibnamefont{Suzuki}},
  \bibinfo{author}{\bibfnamefont{N.}~\bibnamefont{Takemori}},
  \bibinfo{author}{\bibfnamefont{R.}~\bibnamefont{Ishii}},
  \bibinfo{author}{\bibfnamefont{D.}~\bibnamefont{Nishio-Hamane}},
  \bibinfo{author}{\bibfnamefont{R.}~\bibnamefont{Arita}},
  \bibnamefont{et~al.}, \bibinfo{journal}{Nat. Phys.}
  \textbf{\bibinfo{volume}{14}}, \bibinfo{pages}{1119–1124}
  (\bibinfo{year}{2018}).

\bibitem[{\citenamefont{Rothe et~al.}(2012)\citenamefont{Rothe, Hankiewicz,
  Trauzettel, and Guigou}}]{Rothe2012}
\bibinfo{author}{\bibfnamefont{D.~G.} \bibnamefont{Rothe}},
  \bibinfo{author}{\bibfnamefont{E.~M.} \bibnamefont{Hankiewicz}},
  \bibinfo{author}{\bibfnamefont{B.}~\bibnamefont{Trauzettel}},
  \bibnamefont{and} \bibinfo{author}{\bibfnamefont{M.}~\bibnamefont{Guigou}},
  \bibinfo{journal}{Phys. Rev. B} \textbf{\bibinfo{volume}{86}},
  \bibinfo{pages}{165434} (\bibinfo{year}{2012}).

\bibitem[{\citenamefont{Franz and Wiedemann}(1853)}]{Franz:1853WFLaw}
\bibinfo{author}{\bibfnamefont{R.}~\bibnamefont{Franz}} \bibnamefont{and}
  \bibinfo{author}{\bibfnamefont{G.}~\bibnamefont{Wiedemann}},
  \bibinfo{journal}{Ann. Phys.} \textbf{\bibinfo{volume}{165}},
  \bibinfo{pages}{497} (\bibinfo{year}{1853}).

\bibitem[{\citenamefont{Ashcroft and Mermin}(1976)}]{ashcroft1976}
\bibinfo{author}{\bibfnamefont{N.~W.} \bibnamefont{Ashcroft}} \bibnamefont{and}
  \bibinfo{author}{\bibfnamefont{N.~D.} \bibnamefont{Mermin}},
  \emph{\bibinfo{title}{Solid state physics}} (\bibinfo{publisher}{Holt,
  Rinehart and Winston}, \bibinfo{address}{New York [u.a.]},
  \bibinfo{year}{1976}).

\bibitem[{sup()}]{supMat}
\bibinfo{note}{{See Supplemental Material at (insert link) for details on the
  calculations, momentum and energy relaxation, the analytic solutions to the
  linearized Navier-Stokes equations, and complementary figures of the Hall
  response.}}

\bibitem[{\citenamefont{Avron}(1998)}]{avron1998odd}
\bibinfo{author}{\bibfnamefont{J.}~\bibnamefont{Avron}},
  \bibinfo{journal}{Journal of statistical physics}
  \textbf{\bibinfo{volume}{92}}, \bibinfo{pages}{543} (\bibinfo{year}{1998}).

\bibitem[{Foo({\natexlab{b}})}]{Footnote2}
\bibinfo{note}{{Bulk viscous effects are negligible in graphene, at least to
  leading order
  \cite{muller2009graphene,briskot2015collision,Narozhny:2019ehg}.}}

\bibitem[{Foo({\natexlab{c}})}]{Footnote3}
\bibinfo{note}{{We will not approach the charge neutrality point at which the
  imbalance current and the recombination of electrons and holes becomes
  important \cite{Narozhny:2019ehg}.}}

\bibitem[{\citenamefont{Hwang and Das~Sarma}(2008)}]{Hwang:2008MomentumPhonon}
\bibinfo{author}{\bibfnamefont{E.~H.} \bibnamefont{Hwang}} \bibnamefont{and}
  \bibinfo{author}{\bibfnamefont{S.}~\bibnamefont{Das~Sarma}},
  \bibinfo{journal}{Phys. Rev. B} \textbf{\bibinfo{volume}{77}},
  \bibinfo{pages}{115449} (\bibinfo{year}{2008}).

\bibitem[{\citenamefont{Efetov and Kim}(2010)}]{Efetov:PRL:PhononExp}
\bibinfo{author}{\bibfnamefont{D.~K.} \bibnamefont{Efetov}} \bibnamefont{and}
  \bibinfo{author}{\bibfnamefont{P.}~\bibnamefont{Kim}},
  \bibinfo{journal}{Phys. Rev. Lett.} \textbf{\bibinfo{volume}{105}},
  \bibinfo{pages}{256805} (\bibinfo{year}{2010}).

\bibitem[{\citenamefont{Adam et~al.}(2007)\citenamefont{Adam, Hwang, Galitski,
  and Sarma}}]{Shaffique:PNAS:self-consistent}
\bibinfo{author}{\bibfnamefont{S.}~\bibnamefont{Adam}},
  \bibinfo{author}{\bibfnamefont{E.~H.} \bibnamefont{Hwang}},
  \bibinfo{author}{\bibfnamefont{V.~M.} \bibnamefont{Galitski}},
  \bibnamefont{and} \bibinfo{author}{\bibfnamefont{S.~D.} \bibnamefont{Sarma}},
  \bibinfo{journal}{Proc. Natl. Acad. Sci. USA} \textbf{\bibinfo{volume}{104}},
  \bibinfo{pages}{18392} (\bibinfo{year}{2007}).

\bibitem[{\citenamefont{Bistritzer and
  MacDonald}(2009)}]{Cooling:Bistritzer:PRL}
\bibinfo{author}{\bibfnamefont{R.}~\bibnamefont{Bistritzer}} \bibnamefont{and}
  \bibinfo{author}{\bibfnamefont{A.~H.} \bibnamefont{MacDonald}},
  \bibinfo{journal}{Phys. Rev. Lett.} \textbf{\bibinfo{volume}{102}},
  \bibinfo{pages}{206410} (\bibinfo{year}{2009}).

\bibitem[{\citenamefont{Tse and Das~Sarma}(2009)}]{Cooling:Tse:PRB}
\bibinfo{author}{\bibfnamefont{W.-K.} \bibnamefont{Tse}} \bibnamefont{and}
  \bibinfo{author}{\bibfnamefont{S.}~\bibnamefont{Das~Sarma}},
  \bibinfo{journal}{Phys. Rev. B} \textbf{\bibinfo{volume}{79}},
  \bibinfo{pages}{235406} (\bibinfo{year}{2009}).

\bibitem[{\citenamefont{{Narozhny} and
  {Gornyi}}(2021)}]{Narozhny:2021energyrelaxation}
\bibinfo{author}{\bibfnamefont{B.}~\bibnamefont{{Narozhny}}} \bibnamefont{and}
  \bibinfo{author}{\bibfnamefont{I.}~\bibnamefont{{Gornyi}}},
  \bibinfo{journal}{Front. Phys.} \textbf{\bibinfo{volume}{9}},
  \bibinfo{eid}{108} (\bibinfo{year}{2021}).

\bibitem[{\citenamefont{Song et~al.}(2012)\citenamefont{Song, Reizer, and
  Levitov}}]{Disorder-Electron-Phonon}
\bibinfo{author}{\bibfnamefont{J.~C.~W.} \bibnamefont{Song}},
  \bibinfo{author}{\bibfnamefont{M.~Y.} \bibnamefont{Reizer}},
  \bibnamefont{and} \bibinfo{author}{\bibfnamefont{L.~S.}
  \bibnamefont{Levitov}}, \bibinfo{journal}{Phys. Rev. Lett.}
  \textbf{\bibinfo{volume}{109}}, \bibinfo{pages}{106602}
  (\bibinfo{year}{2012}).

\bibitem[{\citenamefont{{Graham} et~al.}(2013)\citenamefont{{Graham}, {Shi},
  {Ralph}, {Park}, and {McEuen}}}]{Graham:2013Supercollisions}
\bibinfo{author}{\bibfnamefont{M.~W.} \bibnamefont{{Graham}}},
  \bibinfo{author}{\bibfnamefont{S.-F.} \bibnamefont{{Shi}}},
  \bibinfo{author}{\bibfnamefont{D.~C.} \bibnamefont{{Ralph}}},
  \bibinfo{author}{\bibfnamefont{J.}~\bibnamefont{{Park}}}, \bibnamefont{and}
  \bibinfo{author}{\bibfnamefont{P.~L.} \bibnamefont{{McEuen}}},
  \bibinfo{journal}{Nat. Phys.} \textbf{\bibinfo{volume}{9}},
  \bibinfo{pages}{103} (\bibinfo{year}{2013}).

\bibitem[{\citenamefont{{Betz} et~al.}(2013)\citenamefont{{Betz}, {Jhang},
  {Pallecchi}, {Ferreira}, {F{\`e}ve}, {Berroir}, and
  {Pla{\c{c}}ais}}}]{Betz:2013NatPh:SupercollisionExperiment}
\bibinfo{author}{\bibfnamefont{A.~C.} \bibnamefont{{Betz}}},
  \bibinfo{author}{\bibfnamefont{S.~H.} \bibnamefont{{Jhang}}},
  \bibinfo{author}{\bibfnamefont{E.}~\bibnamefont{{Pallecchi}}},
  \bibinfo{author}{\bibfnamefont{R.}~\bibnamefont{{Ferreira}}},
  \bibinfo{author}{\bibfnamefont{G.}~\bibnamefont{{F{\`e}ve}}},
  \bibinfo{author}{\bibfnamefont{J.~M.} \bibnamefont{{Berroir}}},
  \bibnamefont{and}
  \bibinfo{author}{\bibfnamefont{B.}~\bibnamefont{{Pla{\c{c}}ais}}},
  \bibinfo{journal}{Nat. Phys.} \textbf{\bibinfo{volume}{9}},
  \bibinfo{pages}{109} (\bibinfo{year}{2013}).

\bibitem[{\citenamefont{Fong et~al.}(2013)\citenamefont{Fong, Wollman, Ravi,
  Chen, Clerk, Shaw, Leduc, and Schwab}}]{Fong:PRX:EnergyRelaxationExperiment}
\bibinfo{author}{\bibfnamefont{K.~C.} \bibnamefont{Fong}},
  \bibinfo{author}{\bibfnamefont{E.~E.} \bibnamefont{Wollman}},
  \bibinfo{author}{\bibfnamefont{H.}~\bibnamefont{Ravi}},
  \bibinfo{author}{\bibfnamefont{W.}~\bibnamefont{Chen}},
  \bibinfo{author}{\bibfnamefont{A.~A.} \bibnamefont{Clerk}},
  \bibinfo{author}{\bibfnamefont{M.~D.} \bibnamefont{Shaw}},
  \bibinfo{author}{\bibfnamefont{H.~G.} \bibnamefont{Leduc}}, \bibnamefont{and}
  \bibinfo{author}{\bibfnamefont{K.~C.} \bibnamefont{Schwab}},
  \bibinfo{journal}{Phys. Rev. X} \textbf{\bibinfo{volume}{3}},
  \bibinfo{pages}{041008} (\bibinfo{year}{2013}).

\bibitem[{\citenamefont{{Narozhny}}(2019)}]{Narozhny:2019ehg}
\bibinfo{author}{\bibfnamefont{B.~N.} \bibnamefont{{Narozhny}}},
  \bibinfo{journal}{Ann. Phys.} \textbf{\bibinfo{volume}{411}},
  \bibinfo{eid}{167979} (\bibinfo{year}{2019}).

\bibitem[{\citenamefont{Landau and Lifshitz}(2013)}]{landau2013fluid}
\bibinfo{author}{\bibfnamefont{L.~D.} \bibnamefont{Landau}} \bibnamefont{and}
  \bibinfo{author}{\bibfnamefont{E.~M.} \bibnamefont{Lifshitz}},
  \emph{\bibinfo{title}{Fluid Mechanics: Course of Theoretical Physics}},
  vol.~\bibinfo{volume}{6} (\bibinfo{publisher}{Elsevier},
  \bibinfo{year}{2013}).

\bibitem[{\citenamefont{Hartnoll et~al.}(2007)\citenamefont{Hartnoll, Kovtun,
  Muller, and Sachdev}}]{Hartnoll:2007Nernst}
\bibinfo{author}{\bibfnamefont{S.~A.} \bibnamefont{Hartnoll}},
  \bibinfo{author}{\bibfnamefont{P.~K.} \bibnamefont{Kovtun}},
  \bibinfo{author}{\bibfnamefont{M.}~\bibnamefont{Muller}}, \bibnamefont{and}
  \bibinfo{author}{\bibfnamefont{S.}~\bibnamefont{Sachdev}},
  \bibinfo{journal}{Phys. Rev. B} \textbf{\bibinfo{volume}{76}},
  \bibinfo{pages}{144502} (\bibinfo{year}{2007}).

\bibitem[{Foo({\natexlab{d}})}]{Footnote4}
\bibinfo{note}{{The nonlocal Coulomb interactions could break Lorentz
  invariance at second order in the velocities \cite{Lucas:2018JPCM}.}}

\bibitem[{\citenamefont{{Narozhny} and {Sch{\"u}tt}}(2019)}]{Narozhny:2019mdg}
\bibinfo{author}{\bibfnamefont{B.~N.} \bibnamefont{{Narozhny}}}
  \bibnamefont{and}
  \bibinfo{author}{\bibfnamefont{M.}~\bibnamefont{{Sch{\"u}tt}}},
  \bibinfo{journal}{\prb} \textbf{\bibinfo{volume}{100}}, \bibinfo{eid}{035125}
  (\bibinfo{year}{2019}).

\bibitem[{\citenamefont{{Fritz} et~al.}(2008)\citenamefont{{Fritz},
  {Schmalian}, {M{\"u}ller}, and {Sachdev}}}]{Fritz:2008qct}
\bibinfo{author}{\bibfnamefont{L.}~\bibnamefont{{Fritz}}},
  \bibinfo{author}{\bibfnamefont{J.}~\bibnamefont{{Schmalian}}},
  \bibinfo{author}{\bibfnamefont{M.}~\bibnamefont{{M{\"u}ller}}},
  \bibnamefont{and}
  \bibinfo{author}{\bibfnamefont{S.}~\bibnamefont{{Sachdev}}},
  \bibinfo{journal}{\prb} \textbf{\bibinfo{volume}{78}}, \bibinfo{eid}{085416}
  (\bibinfo{year}{2008}).

\bibitem[{\citenamefont{Kiselev and Schmalian}(2019)}]{Kiselev:2018slipbdy}
\bibinfo{author}{\bibfnamefont{E.~I.} \bibnamefont{Kiselev}} \bibnamefont{and}
  \bibinfo{author}{\bibfnamefont{J.}~\bibnamefont{Schmalian}},
  \bibinfo{journal}{Phys. Rev. B} \textbf{\bibinfo{volume}{99}}
  (\bibinfo{year}{2019}).

\bibitem[{\citenamefont{Gurzhi}(1963)}]{Gurzhi:1963minimum}
\bibinfo{author}{\bibfnamefont{R.~N.} \bibnamefont{Gurzhi}},
  \bibinfo{journal}{J. Exp. Theor. Phys.} \textbf{\bibinfo{volume}{17}},
  \bibinfo{pages}{521} (\bibinfo{year}{1963}).

\bibitem[{\citenamefont{Gurzhi}(1968)}]{Gurzhi:1968hydrodynamic}
\bibinfo{author}{\bibfnamefont{R.~N.} \bibnamefont{Gurzhi}},
  \bibinfo{journal}{Sov. Phys. Usp.} \textbf{\bibinfo{volume}{11}},
  \bibinfo{pages}{255} (\bibinfo{year}{1968}).

\bibitem[{\citenamefont{Erdmenger et~al.}(2018)\citenamefont{Erdmenger,
  Matthaiakakis, Meyer, and Fern\'andez}}]{PhysRevB.98.195143}
\bibinfo{author}{\bibfnamefont{J.}~\bibnamefont{Erdmenger}},
  \bibinfo{author}{\bibfnamefont{I.}~\bibnamefont{Matthaiakakis}},
  \bibinfo{author}{\bibfnamefont{R.}~\bibnamefont{Meyer}}, \bibnamefont{and}
  \bibinfo{author}{\bibfnamefont{D.~R.} \bibnamefont{Fern\'andez}},
  \bibinfo{journal}{Phys. Rev. B} \textbf{\bibinfo{volume}{98}},
  \bibinfo{pages}{195143} (\bibinfo{year}{2018}).

\bibitem[{\citenamefont{M\"uller and Sachdev}(2008)}]{Muller:PRBcyclotron}
\bibinfo{author}{\bibfnamefont{M.}~\bibnamefont{M\"uller}} \bibnamefont{and}
  \bibinfo{author}{\bibfnamefont{S.}~\bibnamefont{Sachdev}},
  \bibinfo{journal}{Phys. Rev. B} \textbf{\bibinfo{volume}{78}},
  \bibinfo{pages}{115419} (\bibinfo{year}{2008}).

\bibitem[{\citenamefont{Mahajan et~al.}(2013)\citenamefont{Mahajan, Barkeshli,
  and Hartnoll}}]{LorentzRatioNFL}
\bibinfo{author}{\bibfnamefont{R.}~\bibnamefont{Mahajan}},
  \bibinfo{author}{\bibfnamefont{M.}~\bibnamefont{Barkeshli}},
  \bibnamefont{and} \bibinfo{author}{\bibfnamefont{S.~A.}
  \bibnamefont{Hartnoll}}, \bibinfo{journal}{Phys. Rev. B}
  \textbf{\bibinfo{volume}{88}}, \bibinfo{pages}{125107}
  (\bibinfo{year}{2013}).

\bibitem[{\citenamefont{Scaffidi et~al.}(2017)\citenamefont{Scaffidi, Nandi,
  Schmidt, Mackenzie, and Moore}}]{Scaffidi:PRL2017}
\bibinfo{author}{\bibfnamefont{T.}~\bibnamefont{Scaffidi}},
  \bibinfo{author}{\bibfnamefont{N.}~\bibnamefont{Nandi}},
  \bibinfo{author}{\bibfnamefont{B.}~\bibnamefont{Schmidt}},
  \bibinfo{author}{\bibfnamefont{A.~P.} \bibnamefont{Mackenzie}},
  \bibnamefont{and} \bibinfo{author}{\bibfnamefont{J.~E.} \bibnamefont{Moore}},
  \bibinfo{journal}{Phys. Rev. Lett.} \textbf{\bibinfo{volume}{118}},
  \bibinfo{pages}{226601} (\bibinfo{year}{2017}).

\bibitem[{\citenamefont{Mendoza et~al.}(2011)\citenamefont{Mendoza, Herrmann,
  and Succi}}]{mendoza2011preturbulent}
\bibinfo{author}{\bibfnamefont{M.}~\bibnamefont{Mendoza}},
  \bibinfo{author}{\bibfnamefont{H.}~\bibnamefont{Herrmann}}, \bibnamefont{and}
  \bibinfo{author}{\bibfnamefont{S.}~\bibnamefont{Succi}},
  \bibinfo{journal}{Phys. Rev. Lett.} \textbf{\bibinfo{volume}{106}},
  \bibinfo{pages}{156601} (\bibinfo{year}{2011}).

\bibitem[{\citenamefont{Di~Sante et~al.}(2020)\citenamefont{Di~Sante,
  Erdmenger, Greiter, Matthaiakakis, Meyer, Rodr\'\i{}guez~Fern\'andez,
  Thomale, van Loon, and Wehling}}]{DiSante:2019zrd}
\bibinfo{author}{\bibfnamefont{D.}~\bibnamefont{Di~Sante}},
  \bibinfo{author}{\bibfnamefont{J.}~\bibnamefont{Erdmenger}},
  \bibinfo{author}{\bibfnamefont{M.}~\bibnamefont{Greiter}},
  \bibinfo{author}{\bibfnamefont{I.}~\bibnamefont{Matthaiakakis}},
  \bibinfo{author}{\bibfnamefont{R.}~\bibnamefont{Meyer}},
  \bibinfo{author}{\bibfnamefont{D.}~\bibnamefont{Rodr\'\i{}guez~Fern\'andez}},
  \bibinfo{author}{\bibfnamefont{R.}~\bibnamefont{Thomale}},
  \bibinfo{author}{\bibfnamefont{E.}~\bibnamefont{van Loon}}, \bibnamefont{and}
  \bibinfo{author}{\bibfnamefont{T.}~\bibnamefont{Wehling}},
  \bibinfo{journal}{Nat. Commun.} \textbf{\bibinfo{volume}{11}},
  \bibinfo{pages}{3997} (\bibinfo{year}{2020}).

\bibitem[{\citenamefont{B{\"o}ttcher et~al.}(2019)\citenamefont{B{\"o}ttcher,
  Tutschku, Molenkamp, and Hankiewicz}}]{bottcher2019survival}
\bibinfo{author}{\bibfnamefont{J.}~\bibnamefont{B{\"o}ttcher}},
  \bibinfo{author}{\bibfnamefont{C.}~\bibnamefont{Tutschku}},
  \bibinfo{author}{\bibfnamefont{L.~W.} \bibnamefont{Molenkamp}},
  \bibnamefont{and}
  \bibinfo{author}{\bibfnamefont{E.}~\bibnamefont{Hankiewicz}},
  \bibinfo{journal}{Phys. Rev. Lett.} \textbf{\bibinfo{volume}{123}},
  \bibinfo{pages}{226602} (\bibinfo{year}{2019}).

\bibitem[{\citenamefont{Tutschku
  et~al.}(2020{\natexlab{a}})\citenamefont{Tutschku, B{\"o}ttcher, Meyer, and
  Hankiewicz}}]{tutschku2020momentum}
\bibinfo{author}{\bibfnamefont{C.}~\bibnamefont{Tutschku}},
  \bibinfo{author}{\bibfnamefont{J.}~\bibnamefont{B{\"o}ttcher}},
  \bibinfo{author}{\bibfnamefont{R.}~\bibnamefont{Meyer}}, \bibnamefont{and}
  \bibinfo{author}{\bibfnamefont{E.}~\bibnamefont{Hankiewicz}},
  \bibinfo{journal}{Phys. Rev. Research} \textbf{\bibinfo{volume}{2}},
  \bibinfo{pages}{033193} (\bibinfo{year}{2020}{\natexlab{a}}).

\bibitem[{\citenamefont{Tutschku
  et~al.}(2020{\natexlab{b}})\citenamefont{Tutschku, Nogueira, Northe, van~den
  Brink, and Hankiewicz}}]{Tutschku:2020rjq}
\bibinfo{author}{\bibfnamefont{C.}~\bibnamefont{Tutschku}},
  \bibinfo{author}{\bibfnamefont{F.~S.} \bibnamefont{Nogueira}},
  \bibinfo{author}{\bibfnamefont{C.}~\bibnamefont{Northe}},
  \bibinfo{author}{\bibfnamefont{J.}~\bibnamefont{van~den Brink}},
  \bibnamefont{and}
  \bibinfo{author}{\bibfnamefont{E.}~\bibnamefont{Hankiewicz}},
  \bibinfo{journal}{Phys. Rev. B} \textbf{\bibinfo{volume}{102}},
  \bibinfo{pages}{205407} (\bibinfo{year}{2020}{\natexlab{b}}).

\bibitem[{\citenamefont{Moessner et~al.}(2018)\citenamefont{Moessner,
  Sur{\'o}wka, and Witkowski}}]{moessner2018pulsating}
\bibinfo{author}{\bibfnamefont{R.}~\bibnamefont{Moessner}},
  \bibinfo{author}{\bibfnamefont{P.}~\bibnamefont{Sur{\'o}wka}},
  \bibnamefont{and}
  \bibinfo{author}{\bibfnamefont{P.}~\bibnamefont{Witkowski}},
  \bibinfo{journal}{Phys. Rev. B} \textbf{\bibinfo{volume}{97}},
  \bibinfo{pages}{161112} (\bibinfo{year}{2018}).

\bibitem[{\citenamefont{Holder et~al.}(2019)\citenamefont{Holder, Queiroz, and
  Stern}}]{holder2019unified}
\bibinfo{author}{\bibfnamefont{T.}~\bibnamefont{Holder}},
  \bibinfo{author}{\bibfnamefont{R.}~\bibnamefont{Queiroz}}, \bibnamefont{and}
  \bibinfo{author}{\bibfnamefont{A.}~\bibnamefont{Stern}},
  \bibinfo{journal}{Phys. Rev. Lett.} \textbf{\bibinfo{volume}{123}},
  \bibinfo{pages}{106801} (\bibinfo{year}{2019}).

\bibitem[{\citenamefont{Pongsangangan et~al.}(2022)\citenamefont{Pongsangangan,
  Ludwig, Stoof, and Fritz}}]{https://doi.org/10.48550/arxiv.2206.09687}
\bibinfo{author}{\bibfnamefont{K.}~\bibnamefont{Pongsangangan}},
  \bibinfo{author}{\bibfnamefont{T.}~\bibnamefont{Ludwig}},
  \bibinfo{author}{\bibfnamefont{H.~T.~C.} \bibnamefont{Stoof}},
  \bibnamefont{and} \bibinfo{author}{\bibfnamefont{L.}~\bibnamefont{Fritz}},
  \bibinfo{journal}{arXiv:2206.09687}  (\bibinfo{year}{2022}).

\bibitem[{\citenamefont{{Pongsangangan}
  et~al.}(2022)\citenamefont{{Pongsangangan}, {Ludwig}, {Stoof}, and
  {Fritz}}}]{2022arXiv220609694P}
\bibinfo{author}{\bibfnamefont{K.}~\bibnamefont{{Pongsangangan}}},
  \bibinfo{author}{\bibfnamefont{T.}~\bibnamefont{{Ludwig}}},
  \bibinfo{author}{\bibfnamefont{H.~T.~C.} \bibnamefont{{Stoof}}},
  \bibnamefont{and} \bibinfo{author}{\bibfnamefont{L.}~\bibnamefont{{Fritz}}},
  \bibinfo{journal}{arXiv:2206.09694}  (\bibinfo{year}{2022}).

\bibitem[{\citenamefont{Castro~Neto et~al.}(2009)\citenamefont{Castro~Neto,
  Guinea, Peres, Novoselov, and Geim}}]{Castro:ElectronicGraphene}
\bibinfo{author}{\bibfnamefont{A.~H.} \bibnamefont{Castro~Neto}},
  \bibinfo{author}{\bibfnamefont{F.}~\bibnamefont{Guinea}},
  \bibinfo{author}{\bibfnamefont{N.~M.~R.} \bibnamefont{Peres}},
  \bibinfo{author}{\bibfnamefont{K.~S.} \bibnamefont{Novoselov}},
  \bibnamefont{and} \bibinfo{author}{\bibfnamefont{A.~K.} \bibnamefont{Geim}},
  \bibinfo{journal}{Rev. Mod. Phys.} \textbf{\bibinfo{volume}{81}},
  \bibinfo{pages}{109} (\bibinfo{year}{2009}).

\bibitem[{\citenamefont{Novoselov et~al.}(2005)\citenamefont{Novoselov, Geim,
  Morozov, Jiang, Katsnelson, Grigorieva, Dubonos, and
  Firsov}}]{Novoselov:2005:FermionsGraphene}
\bibinfo{author}{\bibfnamefont{K.~S.} \bibnamefont{Novoselov}},
  \bibinfo{author}{\bibfnamefont{A.~K.} \bibnamefont{Geim}},
  \bibinfo{author}{\bibfnamefont{S.~V.} \bibnamefont{Morozov}},
  \bibinfo{author}{\bibfnamefont{D.}~\bibnamefont{Jiang}},
  \bibinfo{author}{\bibfnamefont{M.~I.} \bibnamefont{Katsnelson}},
  \bibinfo{author}{\bibfnamefont{I.}~\bibnamefont{Grigorieva}},
  \bibinfo{author}{\bibfnamefont{S.}~\bibnamefont{Dubonos}}, \bibnamefont{and}
  \bibinfo{author}{\bibfnamefont{a.}~\bibnamefont{Firsov}},
  \bibinfo{journal}{Nature} \textbf{\bibinfo{volume}{438}},
  \bibinfo{pages}{197} (\bibinfo{year}{2005}).

\bibitem[{\citenamefont{{Narozhny} et~al.}(2012)\citenamefont{{Narozhny},
  {Titov}, {Gornyi}, and {Ostrovsky}}}]{Narozhny:2012cdg}
\bibinfo{author}{\bibfnamefont{B.~N.} \bibnamefont{{Narozhny}}},
  \bibinfo{author}{\bibfnamefont{M.}~\bibnamefont{{Titov}}},
  \bibinfo{author}{\bibfnamefont{I.~V.} \bibnamefont{{Gornyi}}},
  \bibnamefont{and} \bibinfo{author}{\bibfnamefont{P.~M.}
  \bibnamefont{{Ostrovsky}}}, \bibinfo{journal}{\prb}
  \textbf{\bibinfo{volume}{85}}, \bibinfo{eid}{195421} (\bibinfo{year}{2012}).

\bibitem[{\citenamefont{{Sheehy} and {Schmalian}}(2007)}]{Sheehy:2007qcs}
\bibinfo{author}{\bibfnamefont{D.~E.} \bibnamefont{{Sheehy}}} \bibnamefont{and}
  \bibinfo{author}{\bibfnamefont{J.}~\bibnamefont{{Schmalian}}},
  \bibinfo{journal}{\prl} \textbf{\bibinfo{volume}{99}}, \bibinfo{eid}{226803}
  (\bibinfo{year}{2007}).

\bibitem[{\citenamefont{Zhang et~al.}(2005)\citenamefont{Zhang, Tan, Stormer,
  and Kim}}]{Zhang:2005QuantumHallExp}
\bibinfo{author}{\bibfnamefont{Y.}~\bibnamefont{Zhang}},
  \bibinfo{author}{\bibfnamefont{Y.-W.} \bibnamefont{Tan}},
  \bibinfo{author}{\bibfnamefont{H.~L.} \bibnamefont{Stormer}},
  \bibnamefont{and} \bibinfo{author}{\bibfnamefont{P.}~\bibnamefont{Kim}},
  \bibinfo{journal}{Nature} \textbf{\bibinfo{volume}{438}},
  \bibinfo{pages}{201} (\bibinfo{year}{2005}).

\bibitem[{\citenamefont{Hwang and Das~Sarma}(2007)}]{Hwang:2007:Dielectric}
\bibinfo{author}{\bibfnamefont{E.~H.} \bibnamefont{Hwang}} \bibnamefont{and}
  \bibinfo{author}{\bibfnamefont{S.}~\bibnamefont{Das~Sarma}},
  \bibinfo{journal}{Phys. Rev. B} \textbf{\bibinfo{volume}{75}},
  \bibinfo{pages}{205418} (\bibinfo{year}{2007}).

\bibitem[{\citenamefont{Strait et~al.}(2011)\citenamefont{Strait, Wang,
  Shivaraman, Shields, Spencer, and Rana}}]{Strait:2011Slow}
\bibinfo{author}{\bibfnamefont{J.~H.} \bibnamefont{Strait}},
  \bibinfo{author}{\bibfnamefont{H.}~\bibnamefont{Wang}},
  \bibinfo{author}{\bibfnamefont{S.}~\bibnamefont{Shivaraman}},
  \bibinfo{author}{\bibfnamefont{V.}~\bibnamefont{Shields}},
  \bibinfo{author}{\bibfnamefont{M.}~\bibnamefont{Spencer}}, \bibnamefont{and}
  \bibinfo{author}{\bibfnamefont{F.}~\bibnamefont{Rana}},
  \bibinfo{journal}{Nano Lett.} \textbf{\bibinfo{volume}{11}},
  \bibinfo{pages}{4902} (\bibinfo{year}{2011}).

\bibitem[{\citenamefont{M{\"u}ller et~al.}(2009)\citenamefont{M{\"u}ller,
  Schmalian, and Fritz}}]{muller2009graphene}
\bibinfo{author}{\bibfnamefont{M.}~\bibnamefont{M{\"u}ller}},
  \bibinfo{author}{\bibfnamefont{J.}~\bibnamefont{Schmalian}},
  \bibnamefont{and} \bibinfo{author}{\bibfnamefont{L.}~\bibnamefont{Fritz}},
  \bibinfo{journal}{Phys. Rev. Lett.} \textbf{\bibinfo{volume}{103}},
  \bibinfo{pages}{025301} (\bibinfo{year}{2009}).

\bibitem[{\citenamefont{Briskot et~al.}(2015)\citenamefont{Briskot, Sch{\"u}tt,
  Gornyi, Titov, Narozhny, and Mirlin}}]{briskot2015collision}
\bibinfo{author}{\bibfnamefont{U.}~\bibnamefont{Briskot}},
  \bibinfo{author}{\bibfnamefont{M.}~\bibnamefont{Sch{\"u}tt}},
  \bibinfo{author}{\bibfnamefont{I.}~\bibnamefont{Gornyi}},
  \bibinfo{author}{\bibfnamefont{M.}~\bibnamefont{Titov}},
  \bibinfo{author}{\bibfnamefont{B.}~\bibnamefont{Narozhny}}, \bibnamefont{and}
  \bibinfo{author}{\bibfnamefont{A.}~\bibnamefont{Mirlin}},
  \bibinfo{journal}{Phys. Rev. B} \textbf{\bibinfo{volume}{92}},
  \bibinfo{pages}{115426} (\bibinfo{year}{2015}).

\bibitem[{Foo({\natexlab{e}})}]{Footnote5}
\bibinfo{note}{{$\eta_H$ has the opposite sign compared to
  Ref.~\cite{Narozhny:2019mdg}.}}

\end{thebibliography}


\begin{widetext}

\section{Supplemental Material}

\subsection{1.~Equation of state for graphene}

Electrons in graphene have a relativistic dispersion relation $\varepsilon_{\bm k}= \pm \hbar v_F |\bm{k}|$ around the two Dirac points at which conduction and valence bands cross in the Brillouin zone \cite{Castro:ElectronicGraphene,Novoselov:2005:FermionsGraphene}, where $v_F\approx 10^6~{\rm m/s}$ is the Fermi velocity. 
The density of states is $\nu(\varepsilon)= N\varepsilon / (2\pi \hbar^2 v_F^2)$, where $N=4$ counts the two spin and two valley degrees of freedom. 
The charge density, imbalance density, and energy density read \cite{Narozhny:2012cdg}
\begin{subequations}
	\begin{align}
		n &= \frac{N k_B^2 T^2}{2\pi \hbar^2 v_F^2R_\Lambda^2}\tilde n, 
		\\
		n_\text{imb} &= \frac{N k_B^2 T^2}{2\pi \hbar^2 v_F^2R_\Lambda^2}\tilde n_\text{imb},
		\\
		n_E &= \frac{N k_B^3 T^3}{\pi \hbar^2 v_F^2 R_\Lambda^2} \tilde n_E,
	\end{align}
\end{subequations}
respectively, where 
\begin{subequations}
	\begin{align}
		\tilde n &=   \text{Li}_2(-e^{-\xi})-\text{Li}_2(-e^{\xi}) , 
		\\
		\tilde n_\text{imb} &= -\text{Li}_2(-e^{-\xi})-\text{Li}_2(-e^{\xi}) ,
		\\
		\tilde n_E &= -\text{Li}_3(-e^{-\xi})-\text{Li}_3(-e^{\xi}) ,
	\end{align}
\end{subequations}
and $\xi=\mu/k_BT$. 
At one-loop, $v_F$ acquires a logarithmic running due to the marginally irrelevant nature of the Coulomb interaction, and the resummation of one-loop diagrams is summarized in the renormalization factor \cite{Narozhny:2019mdg,Sheehy:2007qcs}
\begin{align}
	R_\Lambda=1+\frac{\alpha}{4}\ln\left( \frac{T_\Lambda}{T\max(1,\xi)} \right).
\end{align}
The ultraviolet cutoff scale is set to the temperature $T_\Lambda \approx 8.34 \times 10^4~{\rm K}$ of the band cutoff.
The bare dimensionless Coulomb interaction coupling constant is
\begin{align}
	\alpha=\frac{e^2}{4 \pi  \hbar \epsilon v_F }\approx 0.5 ,
\end{align}
with the dielectric constant $\epsilon\approx4\epsilon_0$ in  graphene on a SiO$_2$ substrate taken from Refs.~\cite{Novoselov:2005:FermionsGraphene,Zhang:2005QuantumHallExp,Hwang:2007:Dielectric}. 

\subsection{2. Two-dimensional charged relativistic hydrodynamics}

We describe the electronic fluid in graphene by two-dimensional relativistic charged hydrodynamics, with the speed of light replaced by the Fermi velocity $v_F$. 
The hydrodynamic equations are the conservation equations of energy, momentum, and charge, \cite{Lucas:2018JPCM,Hartnoll:2007Nernst,Muller:PRBcyclotron}
\begin{align} \label{Eq:Conservation}
	\partial_\nu T^{\nu\mu} &= F^{\mu\nu} J_\nu + \Gamma^\mu, \qquad 
	\partial_\mu J^\mu = 0.
\end{align}
Here, the metric tensor is $\eta_{\mu\nu} = \text{diag}\kc{-,+,+}$.
$T^{\mu\nu}$ and $J^\mu$ are the energy-momentum (EM) tensor and the charge current of the fluid, respectively.
$F^{\mu\nu}$ is the Maxwell tensor of external electromagnetic fields, that also includes the self-consistently determined Vlasov field.
For small non-conservation of energy and momentum, there is a relaxation term $\Gamma^\mu$.
We will express the EM tensor and charge current as functions of the hydrodynamic variables,  velocity $u^\mu=(v_F,\bm v)/{\sqrt{1-|\bm v|/v_F^2}}$,  chemical potential $\mu$, and  temperature $T$. We  work in Landau frame,  
\begin{align}\label{LandauFrame}
	u_\mu T^{\mu\nu} =-n_E u^\nu,\quad
	u_\mu J^\mu = -q nv_F^2\,,
\end{align}
where $n_E$ and $n$ are the energy density and carrier density for  charge $q$ particles, respectively. 
Equation \eqref{LandauFrame} identifies the energy and charge densities in the co-moving frame with the corresponding thermodynamic quantities. The EM tensor and charge current is defined in terms of the hydrodynamic variables to first order in derivatives by \cite{Jensen:2011xb}
\begin{subequations}
	\begin{align}
		T^{\mu\nu} &= n_E u^\mu u^\nu/v_F^2 + P \Delta^{\mu\nu}+\tau^{\mu\nu} , \\
		J^\mu &= q n u^\mu+\nu^{\mu} , 
	\end{align}
\end{subequations}
where $P$ is pressure. 
$\Delta^{\mu\nu}=\eta^{\mu\nu}+u^\mu u^\nu/v_F^2$ is the spatial projector, 
$\tau^{\mu\nu}$ and $\nu^\mu$ are the dissipative as well as non-dissipative first-order corrections. 
Contracting Eq.~\eqref{LandauFrame} with $u^\mu$, we  find the entropy production rate to be 
\begin{align}
	\label{EntropyProduct}
	\partial_\mu J_s^\mu &= \partial_\mu\kc{su^\mu-\frac{\mu_T}{qT}\nu^\mu}
	=
	-\nu^\mu\kc{\partial_\mu\frac{\mu_T}{qT}-\frac1T  F_{\mu\nu}u^\nu}
	-\frac1T\tau^{\mu\nu}\partial_\mu u_\nu\,,
\end{align}
where $J_s^\mu$ is the entropy current, $s$ is the entropy density,  and $V_{\mu \nu }=2\partial _{(\mu }u_{\nu )}-\eta _{\mu \nu } \partial_\rho u^{\rho }$. 
In deriving Eq.~\eqref{EntropyProduct}, we have used the first law of thermodynamics and the Gibbs-Duhem relation
\begin{subequations}
	\begin{align}
		dn_E &= T \, ds+\mu_T \, dn , \\
		n_E + P &= T s + \mu_T n.
	\end{align}
\end{subequations}
By requiring a non-negative entropy production rate and isotropy,  the constitutive relations are obtained to be \cite{avron1998odd,Jensen:2011xb,Lucas:2018JPCM}
\begin{subequations}\label{Eq:Constitutive}
	\begin{align}
		J^{\mu } &= qn u^{\mu } + \sigma_Q \Delta^{\mu \nu }\left(-\frac{T}{q}\partial_\nu\frac{\mu_T }{T}+F_{\nu \rho } u^{\rho }\right),
		\\
		T^{\mu \nu } &= n_E\frac{u^{\mu } u^{\nu }}{v_F^2}+p \Delta^{\mu \nu }-\eta  \Delta^{\mu \rho } \Delta^{\nu \sigma } V_{\rho \sigma } 
		 -\eta _H \epsilon ^{(\mu |\rho \sigma }\frac{u_{\rho }}{v_F}V_{\sigma }{}^{|\nu )}-\zeta  \Delta^{\mu \nu } \partial_\alpha u^{\alpha }.
	\end{align}
\end{subequations}
Here, $\epsilon^{\mu\nu\rho}$ is the totally antisymmetric symbol breaking parity. 
Moreover, $\sigma_Q$ is the quantum critical conductivity, $\zeta$ is the bulk viscosity, $\eta$ is the shear viscosity, and $\eta_H$ is the dissipationless Hall viscosity.
The condition $\sigma_Q,\eta,\zeta\geq0$ is imposed by the non-negativity of the entropy production rate, while the sign of $\eta_H$ is determined by kinetic theory in the next section.
The term proportional to $\eta_H$ breaks both parity and time-reversal symmetry. Inserting Eq.~(1) of the main text and Eq.~\eqref{Eq:Constitutive} into Eq.~\eqref{Eq:Conservation}, we arrive at the equations of motion that are presented in Eq.~(2) of the main text.
\begin{figure}[b!]
	\centering
	\includegraphics[width=0.45\textwidth]{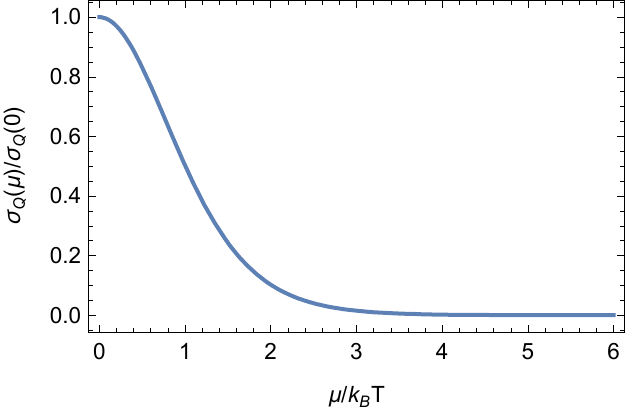}
	\caption{Ratio $\sigma_Q(\mu)/\sigma_Q(\mu=0)$ as a function of $\mu/k_BT$. Figure adapted from Ref.~\cite{Markus:2008quantumcritical}.}
	\label{fig:sigmaQ}
\end{figure}

\subsection{3. Dependence of the transport coefficients on temperature and chemical potential}
The quantum critical conductivity $\sigma_Q$ calculated in Ref.~\cite{Markus:2008quantumcritical} stems from momentum-preserving Coulomb interactions between electrons and holes. 
Its functional dependence on $T$ and $\mu$ is given by
\begin{align}
	\sigma_Q &= \frac{e^2}{\hbar}\frac1{\alpha^2}\frac{N}{2\hat g_1(\mu)}\kc{\tau_{ee}\frac{\alpha^2 k_B T}{\hbar}}^2,
\end{align}
where the electron-electron scattering time reads
\begin{align}
	\tau_{ee}^{-1} = \alpha^2\frac{N}{2\hat g_1(\mu)}\frac{k_BT}{\hbar}  \left[ \frac {N}{2\pi}\ln\kc{2\cosh\left(\frac{\mu}{2k_B T}\right)}  - \frac{(en\hbar v_F)^2}{wT} \right]^{-1},
\end{align}
with the function $\hat g_1(\mu)$ evaluated by the matrix formalism in Ref.~\cite{Markus:2008quantumcritical}. 
As shown in Fig.~\ref{fig:sigmaQ}, $\sigma_Q$ is highly suppressed when approaching the Fermi liquid regime $\mu \gg k_B T$.
In our simulations, we use the viscosities calculated from kinetic theory \cite{Fritz:2008qct,Narozhny:2019mdg,Narozhny:2019ehg} \cite{Footnote5}
\begin{align}
	\zeta\approx 0, 
	\qquad 
	\eta= \frac{k_B^2T^2}{\hbar v_F^2\alpha^2}\tilde\eta,
	\qquad
	\eta_H =\frac{k_B^2T^2}{\hbar v_F^2\alpha^2}\tilde\eta_H,
\end{align}
where
	\begin{subequations}
		\begin{align}
			\tilde\eta
			&=
			\frac{\mathcal T}{4k_B T}
			\begin{pmatrix}
				0 & 0 & 1
			\end{pmatrix}
			\Mh
			\kc{1+\pi^2\gamma_B^2\Teta^{-1}\MK\Teta^{-1}\MK}^{-1}
			\Teta^{-1}
			\begin{pmatrix}
				\tilde n
				\\
				(\xi^2+\pi^2/3)/2
				\\
				3\tilde n_E
			\end{pmatrix},
			\\
			\tilde\eta_H 
			&= -\pi\gamma_B \frac{\mathcal T}{4k_B T}
			\begin{pmatrix}
				0 & 0 & 1
			\end{pmatrix}
			\Mh
			\kc{1+\pi^2\gamma_B^2\Teta^{-1}\MK\Teta^{-1}\MK}^{-1}
			\Teta^{-1}\MK\Teta^{-1}
			\begin{pmatrix}
				\tilde n
				\\
				(\xi^2+\pi^2/3)/2
				\\
				3\tilde n_E
			\end{pmatrix},
		\end{align}
	\end{subequations}
	and
	\begin{align}
		\mathcal T=2k_BT\ln\kc{2\cosh\left( \frac\xi2\right) },
		\qquad
		\gamma_B = \frac{\hbar|e|v_F^2B_z}{\alpha^2k_B^2T^2}.
	\end{align}
	The matrices $\MK$ and $\Mh$ depend on $\xi$ only, while the matrix $\Teta$ depends on $\xi$ and $\alpha$ \cite{Narozhny:2019mdg}. 
	The components of $\Teta$ are
	\begin{align}
		(\Teta)_{ij}
		=\frac{2\pi}{\alpha^2}\frac{\hbar\mathcal T}{N k_B^2 T^2} \tau^{-1}_{ij} 
		=(2\pi)^4
		\int_0^\infty dQ\int_{-\infty}^{+\infty}dW  
		\frac{2\pi Q}{(2\pi)^2}\frac{1}{2\pi}\frac{|\tilde U|^2}{\sinh^2(W)}
		\kd{Y_{00}\tilde Y_{ij}-\tilde Y_{0j}\tilde Y_{0i}},  \label{Integrand}
	\end{align}
	where $q = |{\bm q}|$, and $Q = v_FR_\Lambda\hbar q/2k_BT$ and $W=\hbar\omega/2k_BT$ are the dimensionless momentum and frequency, respectively.
	$Y_{00}$ and $\tilde Y_{ij}$ are functions of $Q$, $W$, and $\xi$ only and their integral expressions are given in Appendix C of Ref.~\cite{Narozhny:2019ehg}. 
	The dynamically screened Coulomb potential is
	\begin{align}
		U(\omega,\bm q) = U_0 \tilde U,
		\quad
		U_0 = \frac{2\pi \hbar \alpha v_F}{ q},
		\quad
		\tilde U = \kc{1+U_0\Pi^R}^{-1}\,,
	\end{align}
	with the polarization operator
	\begin{align}
		\Pi^R
		=\frac{q}{4\pi^2\hbar v_FR_\Lambda^2}\iint_0^1  \frac{dz_1dz_2}{z_1\sqrt{(1-z_1^2)(1-z_2^2)}}
		& \left[(z_1^{-2}-1)\kc{\frac{Q}{z_2Q+W+i\eta}+\frac{Q}{z_2Q-W-i\eta}}J_1(z_1^{-1},z_2,\xi)\right. \nn\\
		&~
		+\left.(1-z_2^{2})\kc{\frac{Q}{z_1^{-1}Q+W+i\eta}+\frac{Q}{z_1^{-1}Q-W-i\eta}}J_2(z_1^{-1},z_2,\xi)\right],
	\end{align}
	with two functions $J_{1,2}(z_1^{-1},z_2,\xi)$ as given in Ref.~\cite{Narozhny:2012cdg}.
	We show the viscosity $\eta$ and Hall viscosity $\eta_H$ as a function of the magnetic field $B_z$ in Fig.~\ref{fig:viscosities}.
	In the Fermi liquid region, the polarization operator at static screening is well approximated by the thermodynamic density of states \cite{Narozhny:2012cdg,Narozhny:2019mdg}, namely, 
	\begin{align}\label{StaticScreening}
		\Pi^R
		\approx \frac{\partial n}{\partial \mu}
		=\frac{N\mathcal T}{2\pi\hbar^2 v_F^2R_\Lambda^2}.
	\end{align}
	\begin{figure}
		\centering
		\includegraphics[width=0.40\linewidth]{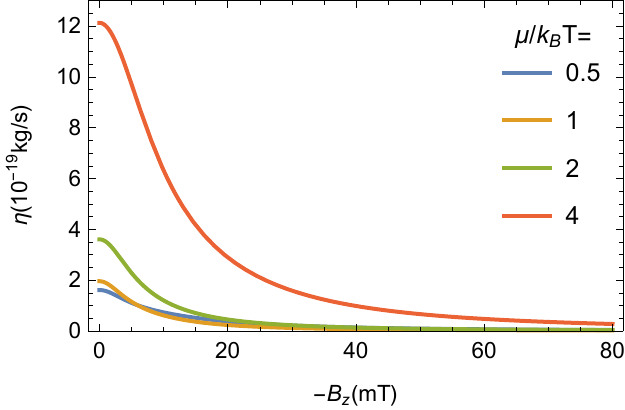}~~~~~~~~
		\includegraphics[width=0.40\linewidth]{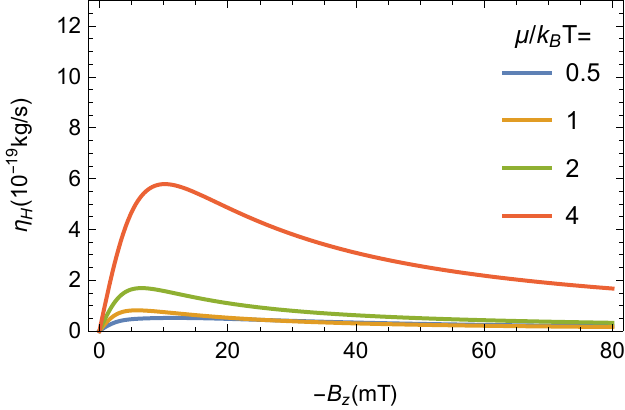}
		\caption{Viscosity $\eta$ and Hall viscosity $\eta_H$ as functions of the magnetic field $B_z$ for different $\mu/k_BT$ at $T=120~{\rm K}$.}
		\label{fig:viscosities}
	\end{figure}

\subsection{4. Momentum relaxation}
Momentum is relaxed in graphene due to electron-impurity  and electron-phonon scattering. In the limit of weak momentum relaxation, the relaxation time approximation can be employed, simplifying the relaxation term to
\begin{align}
	\Gamma^i=-\tau_\text{MR}^{-1}T^{ti}, \qquad i=x,y.
\end{align}
The momentum relaxation time $\tau_\text{MR}$ is related to the electron-impurity scattering time $\tau_\text{imp}^{-1}$ \cite{Markus:2008quantumcritical,Gallagher:2019MontumRelaxation} and electron-phonon scattering time $\tau_\text{ph}^{-1}$ \cite{Hwang:2008MomentumPhonon,Efetov:PRL:PhononExp,Shaffique:PNAS:self-consistent} by Matthiessen’s rule
\begin{align}
	\tau_\text{MR}^{-1}=\tau_\text{imp}^{-1}+\tau_\text{ph}^{-1}.
\end{align}
Here, 
\begin{align}
	\tau_\text{imp}^{-1} = \frac{n_\text{imp}n_\text{imb}}{\hbar w}\kc{\frac{e^2}{4\epsilon}}^2,
	\qquad
	\tau_\text{ph}^{-1}=\frac{ g^2 \mu k_B T}{2\hbar^3v_F^2},
\end{align}
where $n_\text{imp}\approx 2.1 \times 10^9~{\rm cm}^{-2}$ is the density of charged impurities  \cite{Gallagher:2019MontumRelaxation}, $g = D/\sqrt{2\rho_m v_s^2}$ is the electron-phonon coupling, $D\approx 20~{\rm eV}$ is the deformation potential constant, $\rho_m\approx 0.77~{\rm mg/m^2}$ is the mass density of the graphene sheet, and $v_s\approx 2 \times 10^4~{\rm m/s}$ is the longitudinal acoustic (LA) sound velocity. 
We further consider $T>T_\text{BG}$ for $\tau_\text{ph}^{-1}$, where $T_\text{BG} = 2 v_s k_F/k_B$ is the Bloch-Gr{\"u}neisen temperature that is below $100~{\rm K}$ throughout our work. 
Hence, only the acoustic phonons with momentum
$k_\text{ph}\leq 2k_F$ can scatter from electrons. 
For typical scattering times, $\tau_\text{imp}\sim 1~{\rm ps}$ and $\tau_\text{ph}\sim100~{\rm ps}$, so that $\tau_\text{imp}\ll\tau_\text{ph}$ and $\tau_\text{MR}\approx\tau_\text{imp}$.

\subsection{5. Energy relaxation}

The origin of energy relaxation of the electron fluid are electron-phonon collisions involving impurities. 
In electronic cooling experiments in graphene \cite{Betz:2013NatPh:SupercollisionExperiment,Strait:2011Slow,Graham:2013Supercollisions}, the electron and holes heated up by photo excitation or Joule heating relax their energy to phonons.
With the lattice temperature $T_\text{ph}$, the power density of the energy relaxation in the charged fluid at temperature $T$ is found to be \cite{Disorder-Electron-Phonon,Cooling:Bistritzer:PRL,Cooling:Tse:PRB}
\begin{align}\label{CoolingPower}
	\Gamma^0= -A_1(T-T_\text{ph})-A_2(T^3-T_\text{ph}^3)\,,
\end{align}
where 
\begin{subequations}
	\begin{align}
		A_1&= \pi N g^2\nu(\mu)^2\hbar k_F^2 v_s^2 k_B,
		\\
		A_2&= 9.62 \times \frac{g^2\nu^2(\mu)k_B^3}{\hbar k_F l_\text{imp}}.
	\end{align}
\end{subequations}
Here, $l_\text{imp}=v_F\tau_\text{imp}$ is the disorder mean free path. 
The standard cooling pathways mediated by optical and acoustic phonons resulting in the first term is inefficient due to the large value of the optical phonon energy and the strong constraint of the Fermi surface and momentum conservation on the phase space for acoustic phonon scattering  \cite{Cooling:Bistritzer:PRL,Cooling:Tse:PRB}. 
On the other hand, disorder could assists the electron-phonon collisions (supercollisions) by exploring the available phonon phase space \cite{Disorder-Electron-Phonon,Graham:2013Supercollisions,Betz:2013NatPh:SupercollisionExperiment}. It results in the second term and becomes dominated for electron temperatures $T \gtrsim 3T_\text{BG}$.

We will consider that the electron fluid reaches global equilibrium at the lattice temperature, namely $T=T_\text{ph}$.  
For small deviations $\delta T = T - T_\text{ph}$, we can approximate the energy relaxation term
on the right hand side of Eq.~(2a) in the main text at the linear level as
\begin{align}
	\Gamma^0 = -\tau_\text{ER}^{-1}C \, \delta T\,,
\end{align}
with the energy relaxation rate
\begin{align}
	\tau_\text{ER}^{-1}
	=\frac{A_1}{\gamma T}+\frac{3 A_2 T}{\gamma}.
\end{align}
Here, we have used $\delta n_E = C\delta T$, the specific heat $C=\gamma T$, and $\gamma=\frac13\pi^2 N \nu(\mu)k_B^2$. For the typical values of parameters, we have $\tau_\text{ER}\sim 1000{\rm ps}$, which is negligible compared to the other scales in the system.

\subsection{6. Analytic solutions for $\delta\mu$ and $\delta T$}

The analytic solution for $\delta\mu$ and $\delta T$ in the linearized approximation is
\begin{subequations}
	\begin{align}
		\delta \mu &= \frac1{\eta  w} \left[e E_1 n l_G^\text{eff}  \left(e w B_z l_G^2-\mu  \eta _H\right) \csch\left(\frac{W}{2 l_G}\right)   \sinh\left( \frac{y}{l_G}\right) - y B_z \left(e^2 E_1 n w l_G^2+E_2 \eta  \mu  \sigma _Q\right)\right] ,
		\\
		\delta T &= -\frac{T}{\eta  w} \left[ E_2 \eta  y B_z \sigma _Q   +e E_1 n \eta _H l_G^\text{eff} \csch\left(\frac{W}{2 l_G}\right) \sinh \left(\frac{y}{l_G}\right)  \right], 
	\end{align}
\end{subequations}
with $l_G^\text{eff}$ and $E_{1,2}$ as defined in the main text. 
Their profiles, together with $v_x$, are shown in Fig.~\ref{fig:profile2}, from which we see that their boundary values change sign when the magnetic field is increased.
\begin{figure}
	\centering 
	\includegraphics[width = 0.45\textwidth]{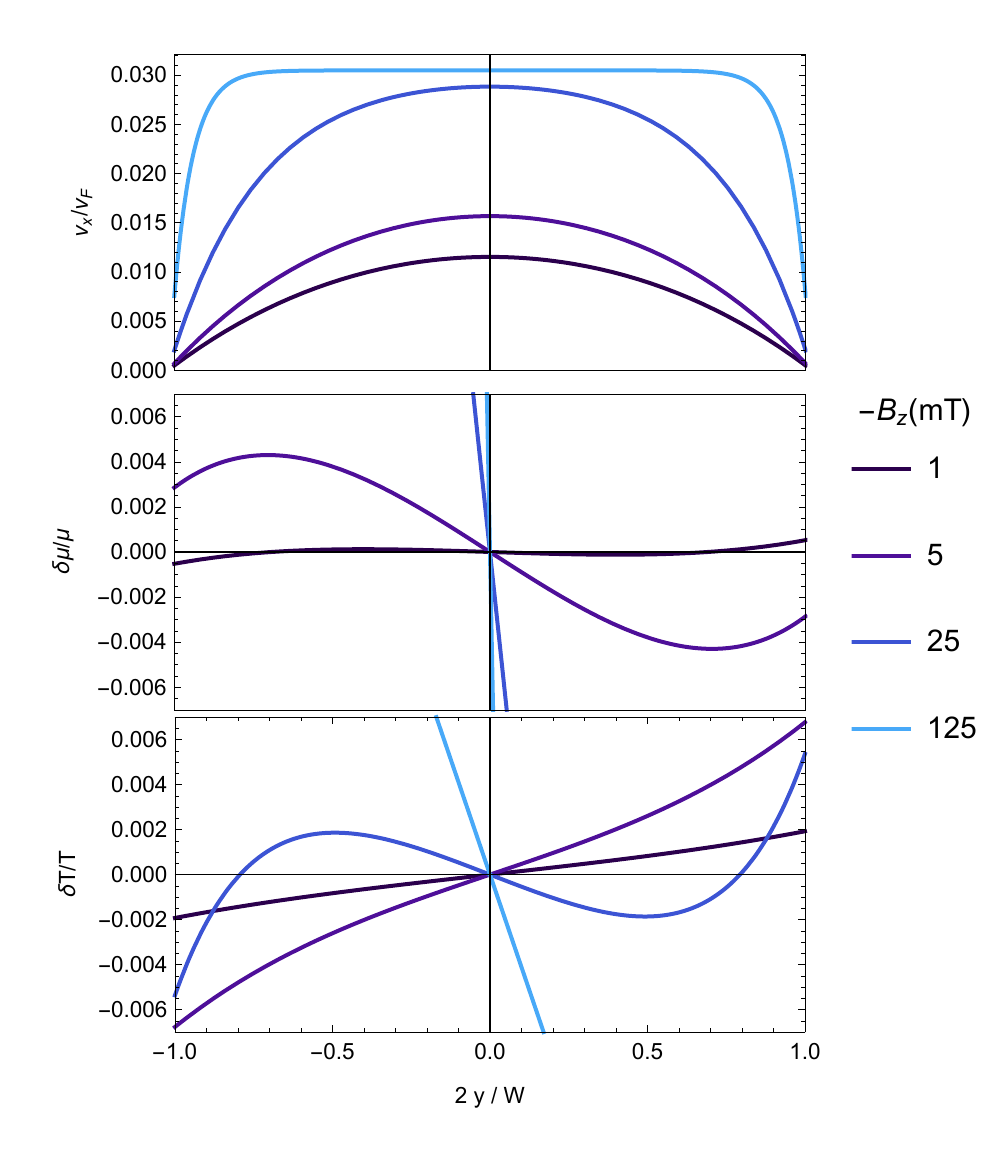}
	\caption{
		The profiles of $v_x$, $\delta\mu$, and $\delta T$ across the channel. We fix $\mu = k_B T$, $T=120~{\rm K}$, $E_x=-1000~{\rm V/m}$, $\nabla_x T = 0$, $W=2~{\rm\mu m}$, $l_s = 0.02~{\rm\mu m}$, and $l_\text{MR} = v_F\tau_\text{MR} = 1.4~{\rm\mu m}$.}
	\label{fig:profile2}
\end{figure}

\subsection{7. Thermoelectric transport in infinite geometries}

We apply a homogeneous electric field $\bm E=E_x\bm e_x$ and temperature gradient $\nabla_x T$ to an infinite sample of graphene, namely,
\begin{align}
	\mu=\mu+xeE_x +\delta\mu(y),\quad 
	T=T+x\nabla_x T+\delta T(y).
\end{align}
The velocity profile is homogeneous and the viscosity terms can be neglected.
The momentum conservation in Eq.~(2) in the main text is simplified as
\begin{align}
	0 &= \partial_x\delta p-BJ_y+\frac{v_x w}{\tau_\text{MR} v_F^2},\\
	0 &= \partial_y\delta p  + BJ_x+\frac{v_y w}{\tau_\text{MR} v_F^2}  - e n \partial_y\phi_V ,
\end{align}
with currents in Eq.~(3) in the main text.
For a system that has infinite width in $x$ and $y$ direction, the general Onsager relations in the DC frequency limit and momentum space are given by \cite{Hartnoll:2007Nernst,Muller:PRBcyclotron}
\begin{align}
	\begin{pmatrix}
		\bm J \\ \bm Q
	\end{pmatrix}
	=
	\begin{pmatrix}
		\bm \sigma & \bm \alpha \\
		{\bm \alpha} T & \bar{\bm \kappa} \\
	\end{pmatrix}
	\begin{pmatrix}
		\bm E \\ 
		-\bm \nabla T  
	\end{pmatrix} .
\end{align}
where $\bm J=(J_x,J_y)$, $\bm Q=(Q_x,Q_y)$, $\bm E=(E_x,E_y)$, $\bm \nabla T=(\nabla_x T,\nabla_y T)$, and $\bm\sigma$, $\bm\alpha$, $\bar{\bm\kappa}$ are $2\times2$ antisymmetric matrices. 
Their matrix components are
\begin{subequations}
	\begin{align}
		\sigma_{xx} &= \sigma_{yy}
		=\frac{\Gamma  \sigma _Q \left(  \gamma +\Gamma+\omega _c^2/\gamma\right)}{(\gamma +\Gamma )^2+\omega _c^2}, 
		\\
		\sigma_{xy}&=-\sigma_{yx}
		=-\frac{\omega _c \sigma _Q \left(  \gamma +2 \Gamma +\omega _c^2/\gamma\right)}{(\gamma +\Gamma )^2+\omega _c^2},
		\\
		\alpha_{xx} &= \alpha_{yy}
		= \Gamma\frac{(\gamma +\Gamma )  \mu \sigma_Q/eT- s\omega_c/B_z}{(\gamma +\Gamma )^2+\omega _c^2}, 
		\\
		\alpha_{xy}&=-\alpha_{yx}
		=\frac{s}{B_z}\frac{ \gamma^2 +\omega _c^2+ \gamma \Gamma (1 -\mu n/sT)}{(\gamma +\Gamma )^2+\omega _c^2}, 
		\\
		\bar\kappa_{xx} &= \bar\kappa_{yy}
		=\frac{\Gamma  \mu ^2 (\gamma +\Gamma ) \sigma _Q/e^2+v_F^2 \left(\gamma  w+\Gamma  s^2 T^2/w\right)}{T \left((\gamma +\Gamma )^2+\omega _c^2\right)}, 
		\\
		\bar{\kappa}_{xy}&=-\bar{\kappa}_{yx}
		=\frac{\omega_c s^2 T^2 - \left(\gamma +(\gamma +2 \Gamma )s T/w\right)\mu  B_z \sigma _Q/e}{T w \left((\gamma +\Gamma )^2+\omega _c^2\right)} , 
	\end{align}
\end{subequations}
where $\Gamma=1/\tau_\text{MR}$,
$\omega_c=e n B_z v_F^2/w$, and $\gamma=B_z^2 v_F^2 \sigma _Q/w$.
However, the boundary conditions in $y$ direction require that $J_y = Q_y = 0$ locally, from which we solve for the two generated sources $E_y$ and $-\nabla_y T$ in terms of the sources $E_x$ and $-\nabla_x T$.
They read
	\begin{align}
		\begin{pmatrix}
			E_y \\ -\nabla_y T
		\end{pmatrix}
		=
		\frac{B_z}{w}\begin{pmatrix}
			-\frac{\mu  \sigma _Q}{e}-e n v_F^2 \tau _\text{MR} & -\frac{\mu ^2 \sigma _Q}{e^2 T}+s v_F^2 \tau _\text{MR} \\
			T \sigma _Q & \frac{\mu  \sigma _Q}{e} 
		\end{pmatrix}
		\begin{pmatrix}
			E_x \\ -\nabla_x T
		\end{pmatrix} .
	\end{align}
	Using these, we obtain the reduced Onsager relation along the $x$ direction
	\begin{align}
		\begin{pmatrix}
			J_x \\
			Q_x
		\end{pmatrix}
		=
		\begin{pmatrix}
			\sigma _Q+ \frac{e^2 n^2 v_F^2 }{w}\tau_\text{MR} & \frac{\mu \sigma _Q}{e T}-\frac{e n s v_F^2 }{w}\tau_\text{MR} \\
			\frac{\mu \sigma _Q}{e}-\frac{e n s T v_F^2 }{w}\tau_\text{MR} & \frac{\mu^2 \sigma _Q}{e^2 T}+\frac{s^2 T v_F^2 }{w}\tau_\text{MR} 
		\end{pmatrix}
		\begin{pmatrix}
			E_x \\ - \nabla_x T
		\end{pmatrix}
	\end{align}
	that takes into account the effects of a finite width in $y$ direction.
	Equation (6) in the main text is of the same form but with $\tau_\text{MR}$ replaced by $\tau_\text{MR}^\text{avg}$. 
	
	\subsection{8. More on the Hall voltage and inverse Nernst effect}
	\begin{figure}
		\centering 
		\includegraphics[width=0.4\textwidth]{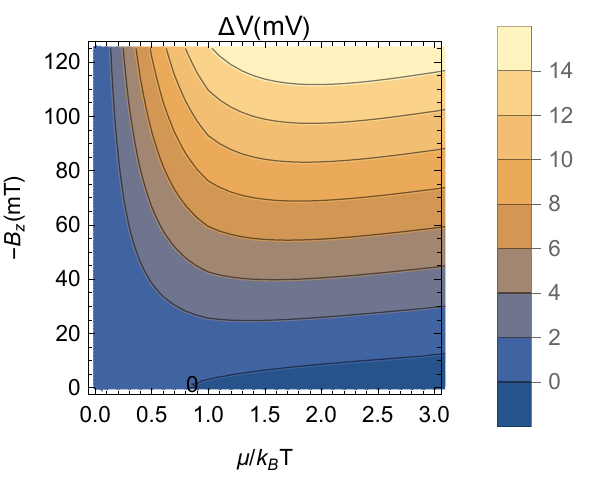}
		\includegraphics[width=0.41\textwidth]{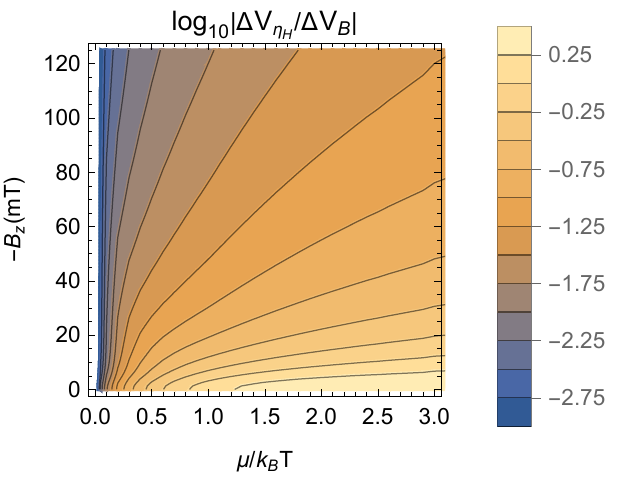}
		\caption{$\Delta V$ and $\log_{10}|\Delta V_{\eta_H}/\Delta V_B|$ as functions of $\mu/k_BT$ and $B_z$ at $T=120~{\rm K}$. The contours of $0$ are labeled. The competition between $\Delta T_B$ and $\Delta T_{\eta_H}$ mainly depends on the magnetic field $B_z$.}
		\label{fig:HVBx}
	\end{figure}
	\begin{figure}
		\centering 
		\includegraphics[width=0.4\textwidth]{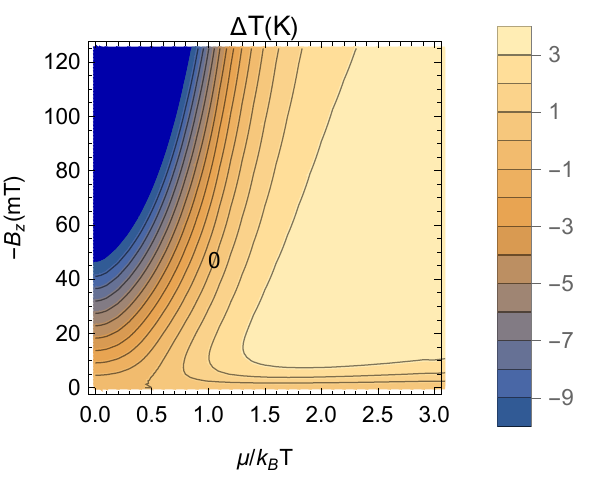}
		\includegraphics[width=0.41\textwidth]{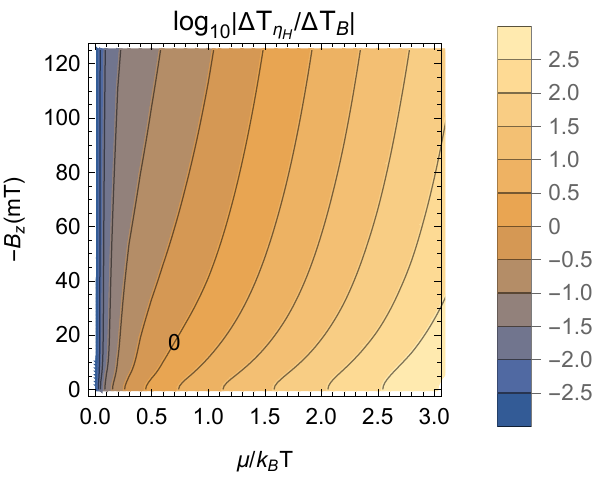}
		\caption{$\Delta T$ and $\log_{10}|\Delta T_{\eta_H}/\Delta T_B|$ as functions of $\mu/k_BT$ and $B_z$ at $T=120~{\rm K}$. The contours of $0$ are labeled. The deep blue regime is clipped as $\Delta T<-10{\rm K}$. The competition between $\Delta T_B$ and $\Delta T_{\eta_H}$ mainly depends on the ratio $\mu/k_BT$ and $\Delta T_{\eta_H}$ becomes dominant when $\mu\gg k_BT$.}
		\label{fig:HTBx}
	\end{figure}
	We show the total Hall voltage $\Delta V$ and temperature gradient $\Delta T$ as functions of $\mu/k_BT$ and magnetic field $B_z$.
	Furthermore, we compare the contributions from Lorentz and Hall viscous effects in Figs.~\ref{fig:HVBx} and \ref{fig:HTBx}. 
	In general, the critical magnetic field for $\Delta V=0$ is small compared to the one for $\Delta T = 0$. 
	It could be explained by comparing the coherent and incoherent contributions in the Lorentz contributions $\Delta V_B$ and $\Delta T_B$, where the coherent and incoherent contributions refer to the parts related to the momentum density $P_x$ sourced by $E_1$ and the quantum critical current density $J_Q$ sourced by $E_2$.
	While the former becomes dominant in the Fermi liquid regime, the latter is dominant in the Dirac regime. 
	In Eq.~(7) in the main text, $\Delta V_B$ contains both coherent and incoherent contributions that stay finite for general values of $\mu/k_BT$.
	In contrast, $\Delta T_B$ in Eq.~(8) in the main text contains the incoherent contribution only, which is suppressed in the Fermi liquid regime. 
	Hence, we find $|\Delta T_{\eta_H}|\gg|\Delta T_B|$ in the Fermi liquid regime, as shown in Fig.~\ref{fig:HTBx}, which allows to extract the pure Hall viscous effect from the total temperature gradient $\Delta T$.
	
\end{widetext}

\end{document}